# Автоматическое доказательство корректности программ с динамической памятью

[1] *Ю.О.Костюков <kostyukov.yurii@gmail.com>*
[1] *К.А.Батоев <konstantin.batoev@gmail.com>*
[1,2] *Д.А.Мордвинов <dmitry.mordvinov@jetbrains.com>*
[1] *М.П.Костицын <mishakosticyn@yandex.ru>*
[1] *А.В.Мисонижник <misonijnik@gmail.com>*
[1] *Санкт-Петербургский государственный университет,*
*199034, Россия, г. Санкт-Петербург, Университетская наб., д. 7/9.*
[2] *JetBrains Research.*

**Аннотация**. В данной работе изучаются теоретические основы автоматической модульной верификации императивных программ с динамической памятью. Вводится модель *композициональной символьной памяти*, и формально доказываются её свойства. Эта модель используется для построения композиционального алгоритма, порождающего *обобщённые кучи*. Они являются термами *исчисления символьных куч*, которые описывают состояния *произвольных* циклических фрагментов программы. Выводимые в этом исчислении *кучи* соответствуют достижимым состояниям исходной программы. В работе также устанавливается соответствие между выводом в этом исчислении и исполнением функциональных программ второго порядка без эффектов.

## *1. Введение*

Бóльшая часть современного программного обеспечения написана на языках с динамически выделяемой памятью, таких как C++, Java, C#. Автоматическая верификация и анализ таких программ является очень трудоёмкой задачей [1]. При этом даже корректные с теоретической точки зрения методы могут оказаться неэффективными из-за больших размеров анализируемых программ [2]. Для решения этой проблемы были предложены *композициональные* техники, которые хорошо зарекомендовали себя на практике [3, 4, 5, 6]. Такие техники выполняют анализ функций в *изоляции*, т. е. вне контекста конкретного вызова, и в дальнейшем переиспользуют промежуточные результаты анализа. Таким образом, можно свести верификацию больших систем к задаче верификации набора небольших фрагментов кода.

Большинство существующих композициональных техник являются *неточными* в том смысле, что они аппроксимируют пространство состояний программы снизу или сверху. Аппроксимирующие снизу подходы рассматривают не все сценарии поведения программы, например, при помощи раскрутки циклов на конечное число шагов [7]. Это позволяет находить ошибки, но не доказывать корректность произвольных программ. Аппроксимирующие сверху подходы анализируют упрощённую версию программы, что на практике приводит к большому числу ложноположительных срабатываний [8]. Таким образом, остаётся актуальной задача точного анализа программ с помощью композициональных техник.

Эту задачу можно решить при помощи введения особой *модели памяти*, представляющей состояния программы. Такая модель должна быть достаточно гибкой и выразительной, чтобы с её помощью можно было описать *произвольные* свойства программы, тем самым охватив все сценарии её поведения. Свойства программ в модели могут быть выражены в виде логических формул.

В данной статье предложен подход к модульной верификации программ с динамической памятью. Подход основывается на новой модели *композициональной символьной памяти*. Эта модель основывается на идее *ленивого инстанцирования* [9] и описывает состояние динамической памяти программы символьными выражениями над значениями входных *ячеек памяти*. Адреса этих ячеек могут быть символьно описаны с использованием содержимого других ячеек; это позволяет выражать эффект любой функции на произвольной рекурсивной структуре данных.

Модель композициональной символьной памяти используется для построения *исчисления символьных куч*, которые, в свою очередь, позволяют описывать поведение произвольной императивной программы с динамической памятью при помощи символьных состояний программы. Термы этого исчисления (т.н. *обобщённые кучи*), построенные по программе, *в точности* описывают эффект этой программы на произвольном состоянии.

В статье также предложена автоматическая процедура композиционального построения обобщённых куч, описывающих произвольные фрагменты императивной программы.

Т.к. для вывода в исчислении символьных куч не требуется хранения контекста, по любой обобщённой символьной куче можно построить эквивалентную ей функциональную программу, сведя задачу проверки корректности императивных программ с динамической памятью к задаче проверки корректности чистых функций. В статье предлагается сведение задачи анализа императивных программ к задаче вывода *уточнённых типов* (refinement types) [10] для функциональных программ, получаемых из обобщённых куч.

Вклад данной работы заключается в следующем: (1) предложена формальная модель *композициональной символьной памяти*; (2) доказаны свойства её корректности; (3) введено *исчисление символьных куч*: выводы в этом исчислении дают все достижимые состояния программы; (4) введено понятие *обобщённых куч*, предложен алгоритм автоматического модульного построения обобщённых куч по императивной программе; (5) предложен подход к сведению задачи поиска вывода в *исчислении символьных куч* к задаче доказательства безопасности функциональных программ.

## 2. Обзор

Существует большое количество работ, посвящённых автоматическому анализу и доказательству корректности программ с динамической памятью [1, 2, 5, 6]. Подходы делятся на *аппроксимирующие снизу*, *аппроксимирующие сверху* и *точные*. Аппроксимирующие снизу подходы исследуют только часть пространства состояний; они пригодны для поиска ошибок в программах, но не годятся для доказательства корректности программ. Подходы, аппроксимирующие пространство состояний сверху, пригодны для доказательства корректности программ, но на практике порождают большое количество ложноположительных срабатываний. Точные подходы исследуют все пространство состояний, но, насколько известно авторам, в данной статье представлен первый *полностью автоматический* подход к точному анализу императивных программ с динамической памятью.

К подходам, аппроксимирующим пространство состояний снизу, можно отнести динамическое символьное исполнение [11] и ограничиваемую проверку моделей [12]. Одна из основных идей в символьном исполнении программ с динамической памятью — ленивое инстанцирование [9]. Оно позволяет производить анализ программ с рекурсивными структурами данных без ручной спецификации размеров этих структур. Существует множество работ, развивающих эту идею [13, 14], но все они следуют идее классического символьного исполнения с раскруткой отношения перехода. Алгоритм, представленный в данной работе, не раскручивает отношение перехода программы, а строит систему ограничений в формализме композициональной символьной памяти, точно описывающих поведения программы.

Как правило, подходы, аппроксимирующие пространство состояний сверху, основаны на *абстрактной интерпретации* программ [2, 6, 8]. Существует множество подходов к абстрактной интерпретации программ с динамической памятью. Подавляющее большинство из них основаны на *логике с разделением* (separation logic) [15]; абстрактный домен таких анализаторов хорошо подходит для автоматического анализа *формы куч* [16], т.е. анализом того, как объекты в куче *связаны* друг с другом, а не их содержимого. Анализаторы, которые пользуются логикой с разделением, являются, как правило, композициональными [2, 5, 6]. Логика с разделением адаптирована для символьного исполнения программ [17]. Основной проблемой логики с разделением является её невыразительность — она хорошо подходит для рассуждения о форме куч, но не подходит для анализа *данных* в динамической памяти, потому что введённая в этой логике разделяющая конъюнкция не позволяет выражать сложных свойств. Существуют подходы на основе логики с разделением, которые позволяют производить более точный анализ динамической памяти вплоть до побайтовых манипуляций с памятью [18]. Однако практическая применимость таких подходов сопряжена с большим количеством ложноположительных срабатываний.

Существуют также работы, основанные на идее сведения задачи верификации программ с динамической памятью к решению системы рекурсивно-логических ограничений [3, 20]. Такие подходы, как и наш, используют SMT-решатели [19] для решения ограничений и вывода индуктивных инвариантов системы. Логические решатели позволяют верифицировать программы, не порождая ложных срабатываний.

## 3. Демонстрационный язык

В данном разделе мы определяем демонстрационный язык программирования с операциями над динамической памятью, для которого позже будет представлена процедура автоматической модульной верификации. Синтаксис языка представлен на рис. 1.

$$
\begin{aligned}
Program &::= Statement^{*} \\
Statement &::= label\colon Statement \\
&\quad \mid Location := Expression \\
&\quad \mid \boldsymbol{goto}\ \{Expression \rightarrow label\}^{+} \\
&\quad \mid \boldsymbol{fail} \mid \boldsymbol{halt} \\
Expression &::= \boldsymbol{null} \mid \boldsymbol{true} \mid \boldsymbol{false} \mid \mathbb{N} \\
&\quad \mid Expression\ BinOp\ Expression \\
&\quad \mid UnOp\ Expression \\
&\quad \mid Location \\
&\quad \mid \boldsymbol{new}\ \{ident\ =\ Expression\}^{+} \\
Location &::= ident \mid Location.ident
\end{aligned}
$$

*Рис. 1. Синтаксис демо-языка.*

*Fig. 1. Demo language syntax*

Состояния программы на этом языке описываются состояниями динамической памяти. Язык не содержит функций, операций ветвления и циклов, однако включает оператор условного перехода по метке $\boldsymbol{goto}\ \{Expression \rightarrow label\}^{+}$. Слева от стрелки находится условие перехода, справа — метка, на которую нужно перейти. Оператор последовательно вычисляет условия слева от стрелок и переходит по первой метке справа от стрелки, чьё условие истинно.

Идентификаторами $ident$ в языке называются идентификаторы в стиле языка C; $BinOp$ и $UnOp$ это простые арифметические и булевы операторы.

Оператор $\boldsymbol{new}\,\{ident = Expression\}^{+}$ выделяет в памяти новый объект с именами полей $ident$, инициализируя каждый из них соответствующим $Expression$. Используя точку, как в определении $Location$, можно получить доступ к полям объектов. Доступ может быть вложенный, например $list.RootNode.Next.Key$.

В каждой переменной находятся либо примитивные значения (**null**, **true**, 42), либо ссылки на объекты в динамической памяти. Оператор $:=$ перезаписывает ссылку. Например, в листинге 1 в строке 3 в переменную $p$ помещается ссылка на $l$, сам объект не копируется. Затем, в строке 7 переменная $p$ начинает ссылаться на новый объект, а переменная $l$ по-прежнему указывает на старый. В строке 10 меняется само содержимое памяти, независимо от переменных $p$ и $l$.

```
1.  RemoveAll:
2.  l := new {Key = x, Next = l}
3.  p := l
4.  RemoveAllIterate:
5.  goto {p.Next = null  -> RemoveAllFinalize,
6.        p.Next.Key = x -> RemoveAllRemoveElement}
7.  p := p.Next
8.  goto {true -> RemoveAllIterate}
9.  RemoveAllRemoveElement:
10. p.Next := p.Next.Next
11. goto {true -> RemoveAllIterate}
12. RemoveAllFinalize:
13. l := l.Next
14.
15. p := l
16. Contains:
17. goto {p = null  -> Exit,
18.       p.Key = x -> Error}
19. p := p.Next
20. goto {true -> Contains}
21.
22. Error: fail
23. Exit: halt
```

*Листинг 1. Программа, модифицирующая динамическую память.*

*Listing 1. Example of heap-manipulating program.*

Далее мы будем рассматривать только корректно типизированные программы: арифметические операции применяются согласно их стандартной семантике; поле и то, что в него записывается, согласованы по типам; выражения в **goto** имеют булевский тип; поля всегда читаются успешно (кроме чтения из **null**); выражения и имена полей, используемые в операторе **new**, согласованы по типам. Также мы предполагаем, что программы не читают из неинициализированных локаций, всякая программа содержит инструкцию **halt**, все метки, на которые есть переходы, определены и имена идентификаторов, меток и ключевые слова языка не пересекаются. Предложенный язык не даёт возможности управлять памятью на низком уровне, в частности, освобождать выделенную память. Таким образом, программы на этом языке допускают всего два вида ошибок, которые необходимо находить: доступ к полю по ссылке, содержащей **null**, и достижимость инструкции **fail**.

## *4. Композициональная символьная память*

В центре нашего подхода стоит формализм композициональных символьных куч. Концепция *композициональной символьной памяти* (КСП), определяемая в этом разделе, основана на идее *ленивого инстанцирования* [9, 13]. Ленивое инстанцирование — это техника, позволяющая строить *конечные* символьные выражения для рекурсивных структур данных, таких как связные списки и деревья. При этом предлагается инициализировать поля рекурсивных структур данных *по требованию* вместо инициализации всей структуры одномоментно, что потребовало бы заранее заданных ограничений на её размер. Это, например, позволяет анализировать списки и деревья, не зная заранее их размер.

```
1. goto {true -> F}
2. G:
3. x := x + 1
4. goto {true -> Exit}
5. F:
6. x := 42
7. goto {true -> G}
8. Exit: halt
```

G (стр. 2-4): $\{x \mapsto LI(x) + 1\}$

F (стр. 5-7): $\{x \mapsto 42\} \circ \{x \mapsto LI(x) + 1\} = \{x \mapsto 42 + 1\}$

*Рис. 2. Пример композиции состояний.*

*Fig. 2. Heap composition example.*

Основная идея КСП состоит в том, чтобы трактовать лениво инициализированные локации $LI(x)$[1] как *ячейки*, в которые будут подставлены значения из контекстного состояния. Например, состояние, описывающее код после метки *G* на рис. 2 (стр. 2-4), содержит незаполненную ячейку *x*. Это означает, что оно описывает *эффект* этого фрагмента кода на *произвольном* контекстном состоянии, т.е. состоянии с произвольным значением *x*. Таким контекстным состоянием может быть, например, состояние после метки *F* до перехода на *G* (стр. 5-6). Чтобы получить полное состояние после метки *F* (стр. 5-7), необходимо заранее подсчитанный эффект *G* применить к текущему состоянию *F*. Для этого нужно заполнить ячейки состояния *G* значениями из текущего контекста *F*. Кучу, полученную в результате этого процесса, мы называем *композицией* куч. Заметим, что адаптация этой идеи к программам произвольной сложности требует некоторых дополнительных усилий.

[1] LI — lazy instantiation

Далее будет описан формализм композициональной символьной памяти. Доказательства теорем приведены в прил. B.

## 4.1 Символьные выражения

Чтобы представлять произвольные состояния программ на нашем демо-языке, необходимо ввести понятие символьных выражений. Определение символьного выражения представлено на рис. 3.

$$\begin{aligned}
term ::= {} & arith \mid loc \\
arith ::= {} & \mathbb{N} \mid arith \pm arith \mid -arith \mid LI^{arith}(loc) \\
& \mid union^{arith}(\langle guard, arith \rangle^*) \\
loc ::= {} & null \mid 0x[0-9]^+ \mid ident \\
& \mid loc.FieldName \mid LI^{loc}(loc) \\
& \mid union^{loc}(\langle guard, loc \rangle^*) \\
guard ::= {} & \top \mid \bot \mid \neg guard \mid guard \land guard \mid guard \lor guard \\
& \mid arith = arith \mid arith < arith \mid loc = loc
\end{aligned}$$

*Рис. 3. Грамматика символьных выражений.*

*Fig. 3. Symbolic terms.*

Символьный терм ($term$) — это либо арифметическое выражение ($arith$), либо символьный адрес локации в памяти ($loc$).

Символьные значения записываются как $LI^*(x)$, что означает «ленивое инстанцирование $x$». Далее всюду тип символьного значения либо не важен, либо очевиден из контекста, потому мы будем его опускать и писать просто $LI(x)$. Заметим, что локация-источник символьного значения $LI$ может быть также символьной, так что, например, термы вида $LI(LI(list).Key) + 1$ также допустимы.

```
1.  a := new {Key = 15, Next = null}
2.  a.Next := a
3.  b.Next := a
4.  c := d.Next
```

$$\begin{cases}
a & \mapsto 0x1 \\
0x1.Key & \mapsto 15 \\
0x1.Next & \mapsto 0x1 \\
LI(b).Next & \mapsto 0x1 \\
c & \mapsto LI(LI(d).Next)
\end{cases}$$

*Рис. 4. Пример различных типов локаций.*

*Fig. 4. Different location types example.*

Локации могут быть *конкретными*, т.е. известными в текущем контексте ($0x[0-9]^+$, порождаются оператором **new**); именованными ($ident$, глобальные переменные, $a, b, c$); ссылками на конкретное поле; ленивыми инстанцированиями других локаций. Например, на рис. 4, где представлен фрагмент кода и соответствующее ему символьное состояние памяти, локация $b$ неизвестна, поэтому запись в её поле $Next$ порождает запись по символьной локации $LI(b).Next$. Аналогично неизвестна локация $d$, но также неизвестен и элемент, на который ссылается её поле $Next$, из-за чего символьная локация $c$ указывает на символьный терм $LI(LI(d).Next)$.

Вернёмся к определению символьных выражений (рис. 3). На этом рисунке $guard$ обозначает *ограничение,* представленное в виде логической формулы (условие пути или условие перехода по метке). Такие формулы «защищают» элементы *символьных объединений*. Символьное объединение[2] ($union$) — это обобщение символа $ite(cond, x, y)$. Как и в случае символьных значений, мы опускаем тип символьных объединений и пишем просто $union(*)$. За символьным объединением мы закрепляем следующую семантику: $x = union(\langle g_1, v_1\rangle, \dots, \langle g_n, v_n\rangle)$ т. и т. т., когда $(g_1 \wedge x = v_1) \vee \dots \vee (g_n \wedge x = v_n)$. Символьные объединения позволяют при помощи одного символьного состояния описать несколько веток исполнения программы.

**Замечание.** Потребуем, чтобы ограничения в объединениях *не пересекались*, т. е. только одно ограничение должно выполняться при подстановке конкретных значений вместо символьных. Однако несколько ограничений могут выполняться одновременно в том случае, если они «защищают» одно и то же значение. Например, допустимы объединения вида

$$union(\langle LI(x) = LI(y), LI(LI(y).Key) + 7\rangle,$$
$$\langle LI(x) = LI(z), LI(LI(z).Key) + 7\rangle).$$

Мы рассматриваем содержимое объединений как множество пар, потому пишем, например, $union(\{\langle g, v\rangle\} \cup X)$. Чтобы избежать перегрузки синтаксиса лишними скобками, мы опускаем их далее при записи одноэлементных множеств. Так, пример выше можно записать следующим образом: $union(\langle g, v\rangle \cup X)$. Также мы опускаем круглые скобки там, где не возникает двусмысленности, например, пишем $union\langle x > 5{,}42\rangle$ или $union\{\langle g_i, v_i\rangle | 1 \leq i \leq n\}$

*Выражение* на рис. 3 — это либо терм, либо ограничение. *Примитивные* выражения — это натуральные числа, именованные локации, конкретные адреса в памяти, $\boldsymbol{null}$, $\top$ и $\bot$. *Операциями* являются сложение, вычитание, унарный минус, сравнения, логические связки и чтение поля. В тех случаях, когда вид операции не важен, мы пользуемся следующей нотацией: $op(e_1, \dots, e_n)$.

**Замечание.** Равенство символьных термов является *семантическим*, например, $2 * (x + 1) = x + x + 4 - 2$ и $union(\langle x + 5 = y + 4{,}7\rangle, \langle \bot, 42\rangle) = union\langle x + 1 = y, 7\rangle$.

Далее мы перечисляем некоторые очевидные свойства символьного объединения.

**Утверждение 1.**

(a) $union\langle \top, v\rangle = v$

[2] Понятие символьных объединений заимствовано из [21]

(b) $union(\langle \bot, v\rangle \cup X) = union(X)$

$$union(\langle g, union(\langle g_1, v_1\rangle, \dots, \langle g_n, v_n\rangle)\rangle \cup X) = \\ = union(\{\langle g \wedge g_1, v_1\rangle, \dots, \langle g \wedge g_n, v_n\rangle\} \cup X)$$

В частности,

- $union(\langle g, union(\varnothing)\rangle \cup X) = union(X)$
- $union\langle g_1 \vee \dots \vee g_n, union(\langle g_1, v_1\rangle, \dots, \langle g_n, v_n\rangle)\rangle = \\ = union(\langle g_1, v_1\rangle, \dots, \langle g_n, v_n\rangle)$

(c) $union(\{\langle g_1, v\rangle, \dots, \langle g_n, v\rangle\} \cup X) = union(\langle g_1 \vee \dots \vee g_n, v\rangle \cup X)$

В частности,

$$union(\{\langle g, v\rangle, \langle g \wedge g_1, v\rangle, \dots, \langle g \wedge g_n, v\rangle\} \cup X) = union(\langle g, v\rangle \cup X)$$

(d) $op(union(\langle g_1^1, e_1^1\rangle, \dots, \langle g_{n_1}^1, e_{n_1}^1\rangle), \dots, union(\langle g_1^m, e_1^m\rangle, \dots, \langle g_{nm}^m, e_{nm}^m\rangle)) = \\ = union(\{\langle g_{i_1}^1 \wedge \dots \wedge g_{i_m}^m, op(e_{i_1}^1, \dots, e_{i_m}^m)\rangle | 1 \le i_j \le n_j\})$

В частности, для непересекающихся ограничений $g_1, \dots, g_n$,

$$op(union(\langle g_1, e_1^1\rangle, \dots, \langle g_n, e_n^1\rangle), \dots, union(\langle g_1, e_1^m\rangle, \dots, \langle g_n, e_n^m\rangle)) = \\ = union(\langle g_1, op(e_1^1, \dots, e_1^m)\rangle, \dots, \langle g_n, op(e_n^1, \dots, e_n^m)\rangle)$$

## 4.2 Символьные кучи

**Определение 1.** *Символьная куча* — это частичная функция $\sigma: loc \to term$, удовлетворяющая следующему требованию *(инвариант кучи):*

$$\forall x, y \in dom(\sigma), union\langle x = y, \sigma(x)\rangle = union\langle x = y, \sigma(y)\rangle. \qquad (1)$$

Заметим, что это более сильное ограничение, чем накладывает само понятие функции ($x \equiv y \Rightarrow \sigma(x) \equiv \sigma(y)$). Это связано с тем, что равенство термов и локаций в нашем подходе является не синтаксическим, а семантическим. Так, например, функция $\{LI(x).Key \mapsto 10;\ LI(y).Key \mapsto 15\}$ не является символьной кучей, т.к. при подстановке вместо $x$ и $y$, например, $0x1$, получится, что этот адрес указывает на два разных значения. Напротив, следующая функция является символьной кучей: $\{LI(x).Key \mapsto union(\langle LI(x) = LI(y), 15\rangle, \langle LI(x) \ne LI(y), 10\rangle); LI(y).Key \mapsto 15\}$.

**Определение 2.** Пустая куча $\epsilon$ — это частичная функция с областью определения $dom(\epsilon) = \varnothing$ (она, очевидно, удовлетворяет (1)).

**Определение 3** Пусть $x \in dom(\sigma)$ или $x$ — символьная локация. Тогда определим чтение локации $x$ в символьной куче $\sigma$ следующим образом:

$$read(\sigma, x) \stackrel{\text{def}}{=} union(\{\langle x = l, \sigma(l)\rangle | l \in dom(\sigma)\} \cup \langle \bigwedge_{l \in dom(\sigma)} x \ne l, LI(x)\rangle). \qquad (2)$$

Интуитивно, чтение пытается сопоставить ссылку $x$ (возможно, символьную) с каждым адресом локации в $\sigma$ (также, возможно символьным). Если ссылка и некоторый адрес совпали, то результатом чтения будет значение, лежащее по этому адресу. Если не было найдено ни одного совпадения, то возвращается символьное значение $LI(x)$.

Очевидно, что при $x \in dom(\sigma)$ выполнено $read(\sigma, x) = \sigma(x)$. Одно из ограничений $x = l$ будет выполняться, тогда как один из элементов конъюнкции $\bigwedge_{l \in dom(\sigma)} x \neq l$ будет, наоборот, невыполним, следовательно $read(\sigma, x)$ не сможет вернуть значение $LI(x)$.

Сделаем следующее наблюдение: фактически, из опр. 3 следует, что множество ограничений в формуле (2) может содержать пересечения, т.е. в куче могут содержаться две (или более) символьные локации, которые совпадают при некоторых конкретных подстановках. Инвариант символьной кучи (1) позволяет обойти возможную проблему с совпадающими адресами: благодаря ему при совпадении ограничений не будет конфликтов между «защищаемыми» значениями.

**Пример 1.** *Пусть* $\sigma = \{0x1.A \mapsto 42; LI(x).B \mapsto union(\langle LI(x).B = 0x1.A, 42\rangle, \langle LI(x).B \neq 0x1.A, 7\rangle)\}$. *Тогда*

$$read(\sigma, 0x1.A) = union(\langle 0x1.A = 0x1.A, 42\rangle, \langle 0x1.A = LI(x).B, union(\dots)\rangle, \langle 0x1.A \neq 0x1.A \land LI(x).B \neq 0x1.A, LI(0x1.A)\rangle) =$$

$$= union(\langle \top, 42\rangle, \langle 0x1.A = LI(x).B, union(\dots)\rangle, \langle \bot, LI(0x1.A)\rangle) = 42$$

$$read(\sigma, LI(y).B) =$$

$$= union(\langle LI(y).B = 0x1.A, 42\rangle, \langle LI(y).B = LI(x).B, union(\langle LI(x).B = 0x1.A, 42\rangle, \langle LI(x).B \neq 0x1.A, 7\rangle)\rangle, \langle LI(y).B \neq 0x1.A \land LI(y).B \neq LI(x).B, LI(LI(y).B)\rangle) \overset{\text{утв. 1}}{=}$$

$$= union(\langle LI(y).B = 0x1.A, 42\rangle, \langle LI(y).B = LI(x).B \land LI(x).B \neq 0x1.A, 7\rangle, \langle LI(y).B \neq 0x1.A \land LI(y).B \neq LI(x).B, LI(LI(y).B)\rangle)$$

## 4.3 Композиция символьных куч

**Определение 4.** *Уточнение* выражения $e$ в контексте символьной кучи $\sigma$ обозначим $\sigma \bullet e$ и определим следующим образом.

1. Если $e$ — это примитивное значение, то $\sigma \bullet e \overset{\text{def}}{=} e$.
2. $\sigma \bullet op(e_1, \dots, e_n) \overset{\text{def}}{=} op(\sigma \bullet e_1, \dots, \sigma \bullet e_n)$.
3. $\sigma \bullet union\{\langle g_1, t_1\rangle, \dots, \langle g_n, t_n\rangle\} \overset{\text{def}}{=} union\{\langle \sigma \bullet g_1, \sigma \bullet t_1\rangle, \dots, \langle \sigma \bullet g_n, \sigma \bullet t_n\rangle\}$.
4. $\sigma \bullet LI(l) \overset{\text{def}}{=} read(\sigma, \sigma \bullet l)$.

Интуитивно, $\sigma \bullet e$ — это выражение, получаемое подстановками значений из $\sigma$ в символьные ячейки $e$: первые три пункта определения сохраняют структуру $e$, а п.4 заполняет ячейку значением из $\sigma$.

**Определение 5.** *Композиция символьных куч* $\sigma$ *и* $\sigma'$ — это частичная функция $\sigma \circ \sigma': loc \to term$, определяемая так:

$$(\sigma \circ \sigma')(x) \overset{\text{def}}{=}$$

$$\overset{\text{def}}{=} union(\{\langle x = \sigma \bullet l, \sigma \bullet (\sigma'(l))\rangle \mid l \in dom(\sigma')\} \cup \langle \bigwedge_{l \in dom(\sigma')} x \neq \sigma \bullet l, \sigma(x)\rangle).$$

Композиция $\sigma \circ \sigma'$ определена на всех локациях, удовлетворяющих ограничению $x = \sigma \bullet l$, где $l \in dom(\sigma')$, и на всех локациях $dom(\sigma)$ (здесь и далее запись $\{\sigma \bullet a | a \in A\}$ сокращается как $\sigma \bullet A$). Из этого следует, что:

$$dom(\sigma \circ \sigma') = dom(\sigma) \cup \sigma \bullet dom(\sigma'). \qquad (3)$$

Композиция символьных куч отражает последовательную композицию в программировании: если $\sigma_1$ — это эффект фрагмента кода A и $\sigma_2$ — эффект фрагмента кода B, тогда $\sigma_1 \circ \sigma_2$ — это эффект A;B. Интуитивно, $\sigma_1 \circ \sigma_2$ — это символьная куча, полученная заполнением символьных ячеек из $\sigma_2$ значениями из контекста $\sigma_1$ с последующей их записью в контекст $\sigma_1$.

**Пример 2.** Пусть $\sigma = \{x \mapsto 42; y \mapsto 7\}$ и $\sigma' = \{y \mapsto LI(x) - LI(y)\}$. Тогда $\sigma \circ \sigma' = \{x \mapsto 42; y \mapsto 42 - 7\}$.

**Теорема 1.** Если $\sigma$ и $\sigma'$ — символьные кучи, то $\sigma \circ \sigma'$ также символьная куча.

**Теорема 2.** Для произвольной символьной кучи $\sigma$, локации $x$ и выражения $e$ справедливо следующее:

(a) $read(\epsilon, x) = LI(x)$

(b) $\epsilon \bullet e = e$

(c) $\epsilon \circ \sigma = \sigma$

(d) $\sigma \circ \epsilon = \sigma$

Стоит отметить, что имеется некоторое сходство между чтением (опр. 3) и композицией (опр. 5): объединения осуществляют поиск $x$ среди локаций кучи. Если поиск был успешен, возвращается соответствующее (возможно изменённое) значение, в ином случае — значение по умолчанию. Воспользуемся этим сходством для определения оператора $find$, который далее используется для трансляции обобщённых куч в чистые функции.

**Определение 6.**

$$find(\sigma, x, \tau, d) \stackrel{\text{def}}{=}$$
$$\stackrel{\text{def}}{=} union(\{\langle x = \tau \bullet l, \tau \bullet (\sigma(l))\rangle | l \in dom(\sigma)\} \cup \langle \bigwedge_{l \in dom(\sigma)} x \neq \tau \bullet l, d\rangle)$$

Теор. 2 позволяет компактно выразить $read$ и композицию куч через $find$:

$$read(\sigma, x) = find(\sigma, x, \epsilon, LI(x)) \qquad (4)$$
$$(\sigma \circ \sigma')(x) = find(\sigma', x, \sigma, \sigma(x)) \qquad (5)$$

**Теорема 3.** Для всех символьных куч $\sigma$, $\sigma'$ и символьных локаций $x$ справедливо следующее:

$$\sigma \bullet read(\sigma', x) = read(\sigma \circ \sigma', \sigma \bullet x).$$

Теорема 3 говорит о корректности чтения из композиции состояний. Например, имея состояния $\sigma_F$ и $\sigma_G$ исполняемых друг за другом фрагментов кода (как на рис. 5) и читая после фрагмента $G$ переменную $x$, можно прочитать переменную из позднего состояния $G$, а затем заполнить результат из контекста $F$. Однако возможно также прежде воспроизвести эффект $G$ поверх эффекта $F$ ($\sigma_F \circ \sigma_G$), и прочитать $x$ из уточнённого состояния. Теорема 3 утверждает, что результаты двух таких операций совпадут.

```
goto {true -> F}
G:
x := a.Key + 5
goto {true -> Exit}
F:
a := new {Key = 10}
goto {true -> G}
Exit:
r := x
halt
```

$$\sigma_G = \{x \mapsto (LI(LI(a).Key) + 5)\}$$
$$read(\sigma_G, x) = \sigma_G(x) = LI(LI(a).Key) + 5$$
$$\sigma_F \bullet read(\sigma_G, x) = 10 + 5$$

$$\sigma_F \circ \sigma_G = \{a \mapsto 0x1;\ 0x1.Key \mapsto 10;\ x \mapsto (10 + 5)\}$$
$$read(\sigma_F \circ \sigma_G, \sigma_F \bullet x) = read(\sigma_F \circ \sigma_G, x) = 10 + 5$$

*Рис. 5. Чтение из композиции состояний.*

*Fig. 5. Read from composition example.*

**Теорема 4.** Для всех символьных куч $\sigma$, $\sigma'$ и символьного выражения $e$ справедливо следующее: $(\sigma \circ \sigma') \bullet e = \sigma \bullet (\sigma' \bullet e)$.

Допустим, имеется три последовательных фрагмента кода с метками $F$, $G$ и $H$, причём каждый фрагмент заканчивается переходом на следующую метку в этом списке. Интуитивно, итоговое символьное состояние всего кода не должно зависеть от порядка применения эффектов этих фрагментов. Следующая теорема показывает, что КСП обладает указанным свойством.

**Теорема 5.** Для всех символьных куч $\sigma_1$, $\sigma_2$ и $\sigma_3$ справедливо следующее:

$$(\sigma_1 \circ \sigma_2) \circ \sigma_3 = \sigma_1 \circ (\sigma_2 \circ \sigma_3).$$

**Теорема 6.** Пусть $\Sigma$ — множество всех символьных куч. Тогда $(\Sigma, \circ)$ — моноид.

*Доказательство.* Из теоремы 1 следует замкнутость множества $\Sigma$ относительно операции $\circ$, по теореме 2 $\epsilon$ — нейтральный элемент, и по теореме 5, операция $\circ$ ассоциативна.

## 4.4 Объединение символьных куч

До сих пор мы рассматривали символьные состояния только для линейных фрагментов кода, в которых все переходы по меткам были безусловными ($\boldsymbol{goto}\,\{true \to \cdots\}$). Для более сложных состояний необходим новый оператор, позволяющий объединить несколько состояний в одно.

**Определение 7.** *Объединением* $\sigma = merge(\langle g_1, \sigma_1\rangle, \dots, \langle g_n, \sigma_n\rangle)$ символьных куч $\sigma_1, \dots, \sigma_n$ по непересекающимся ограничениям $g_1, \dots, g_n$ будем называть частичную функцию с $dom(\sigma) = \bigcup_{i=1}^{n} dom(\sigma_i)$, для которой выполняется следующее: $(merge\langle g_i, \sigma_i\rangle)(x) \stackrel{\text{def}}{=} union\langle g_i, read(\sigma_i, x)\rangle$.

**Теорема 7.** Для любой символьной кучи $\sigma$ и произвольных символьных локаций $x$, $y$ справедливо следующее:

$$union(\langle x = y, read(\sigma, x)\rangle) = union(\langle x = y, read(\sigma, y)\rangle).$$

Далее покажем, что оператор $merge$ обладает интуитивными свойствами. **Теорема 8.** Для любых символьных куч $\sigma_1, \dots, \sigma_n$ и произвольных непересекающихся ограничений $g_1, \dots, g_n$, справедливо утверждение о том, что $merge\langle g_i, \sigma_i \rangle$ — символьная куча.

**Теорема 9.** Для любых символьных куч $\sigma_1, \dots, \sigma_n$, произвольных непересекающихся ограничений $g_1, \dots, g_n$ и для любых локаций $x$ справедливо следующее:

$$read(merge\langle g_i, \sigma_i \rangle, x) = union\langle g_i, read(\sigma_i, x) \rangle.$$

Как уже было показано, результат вычисления символьного состояния не зависит от порядка применения эффектов. Необходимо показать, что это свойство КСП выполняется и для нового оператора объединения состояний. Для этого нужно рассмотреть два случая расстановки объединения и композиции, представленные на рис. 6.

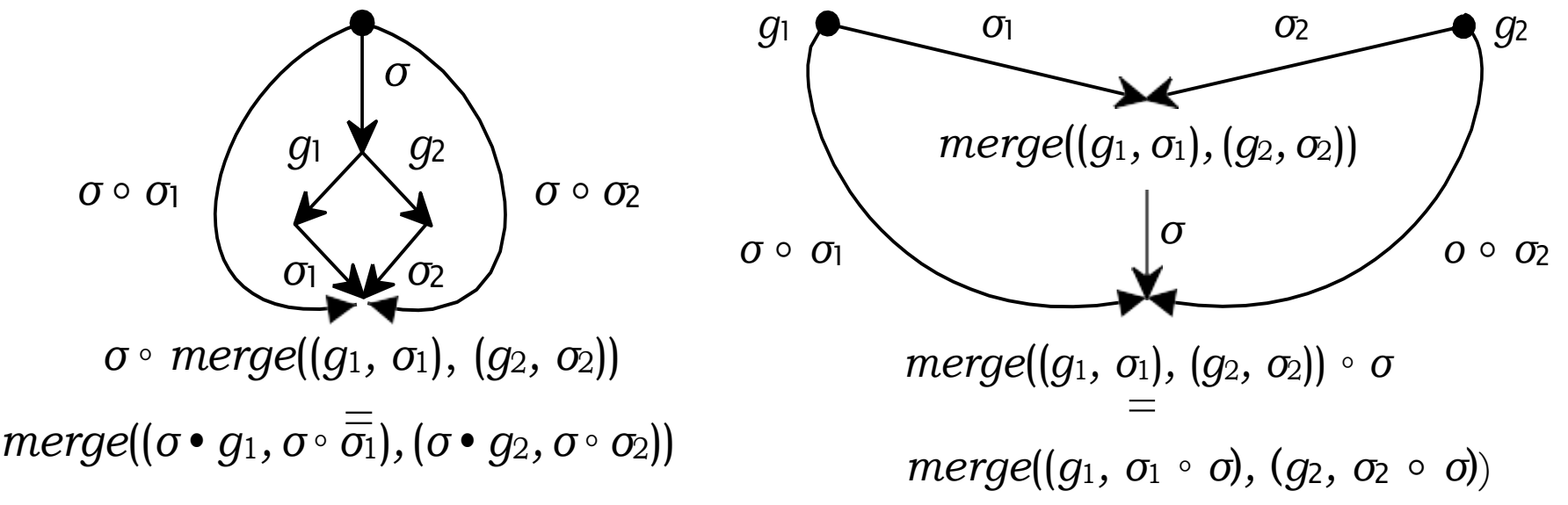

*Рис. 6. Композиция объединения куч.*

*Fig. 6. Composition of merge.*

**Теорема 10.** Для любых символьных куч $\sigma, \sigma_1, \dots, \sigma_n$ и произвольных непересекающихся ограничений $g_1, \dots, g_n$ выполняется следующее утверждение:

$$\sigma \circ merge(\langle g_1, \sigma_1 \rangle, \dots, \langle g_n, \sigma_n \rangle) = merge(\langle \sigma \bullet g_1, \sigma \circ \sigma_1 \rangle, \dots, \langle \sigma \bullet g_n, \sigma \circ \sigma_n \rangle).$$

Интересно, что симметричный случай гораздо сложнее. Например, рассмотрим $\sigma_1 = \{x \mapsto LI(a)\}$, $\sigma_2 = \{x \mapsto LI(b)\}$, $\sigma = \{LI(x).Key \mapsto 42\}$. Тогда:

$$dom(merge(\langle g, \sigma_1 \rangle, \langle \neg g, \sigma_2 \rangle) \circ \sigma) =$$
$$= dom(merge(\langle g, \sigma_1 \rangle, \langle \neg g, \sigma_2 \rangle)) \cup merge(\langle g, \sigma_1 \rangle, \langle \neg g, \sigma_2 \rangle) \bullet dom(\sigma) =$$
$$= \{x, union(\langle g, LI(a).Key \rangle, \langle \neg g, LI(b).Key \rangle)\},$$

что не то же самое, что

$$dom(merge(\langle g, \sigma_1 \circ \sigma \rangle, \langle \neg g, \sigma_2 \circ \sigma \rangle)) = dom(\sigma_1 \circ \sigma) \cup dom(\sigma_2 \circ \sigma) =$$
$$= \{LI(a).Key, LI(b).Key, x\}.$$

Чтобы избежать проблем такого вида, далее мы будем требовать, чтобы символьные ячейки $LI(*)$ удовлетворяли следующему дополнительному свойству: для любых непересекающихся ограничений $g_1, \dots, g_n$ и символьных локаций $x_1, \dots, x_n$ должно выполняться:

$$LI(union(\langle g_1, x_1 \rangle, \dots, \langle g_n, x_n \rangle)) = union(\langle g_1, LI(x_1) \rangle, \dots, \langle g_n, LI(x_n) \rangle). \quad (6)$$

**Теорема 11.** Для любой символьной кучи $\sigma$ и произвольных непересекающихся ограничений $g_1, \dots, g_n$ , а также любых локаций $x_1, \dots, x_n$ справедливо следующее утверждение:

$$read(\sigma, union\langle g_i, x_i\rangle) = union\langle g_i, read(\sigma, x_i)\rangle.$$

Теперь необходимо сформулировать вспомогательное утверждение, которое позволит доказать симметричную теорему о композиции с объединением: $merge\langle g_i, \sigma_i\rangle \bullet e = union\langle g_i, \sigma_i \bullet e\rangle$. Однако оно не всегда верно: может случиться так, что $g_1 \vee \dots \vee g_n \neq \top$, что, в свою очередь, может привести к попытке уточнить терм в несуществующей куче. Например, можно ожидать, что уточнение терма 42 в любой куче даст 42, однако в рамках наших определений мы получим $union(\langle g_1, 42\rangle, \dots, \langle g_1, 42\rangle) = union\langle g_1 \vee \dots \vee g_n, 42\rangle \neq 42$. Чтобы указанное выше утверждение выполнялось, необходимо ограничить результат условием существования вычисления $g_1 \vee \dots \vee g_n$.

**Теорема 12.** Для любых символьных куч $\sigma_1, \dots, \sigma_n$, произвольных непересекающихся ограничений $g_1, \dots, g_n$ и некоторого выражения $e$ справедливо следующее утверждение:

$$union\langle g_1 \vee \dots \vee g_n, merge\langle g_i, \sigma_i\rangle \bullet e\rangle = union\langle g_i, \sigma_i \bullet e\rangle.$$

Таким образом, нельзя приравнять $merge(\langle g_1, \sigma_1\rangle, \dots, \langle g_n, \sigma_n\rangle) \circ \sigma$ и $merge(\langle g_1, \sigma_1 \circ \sigma\rangle, \dots, \langle g_n, \sigma_n \circ \sigma\rangle)$ как теоретико-множественные объекты. Однако следующая теорема показывает, что определение операции чтения позволяет избежать этой проблемы: в теоретико-множественном смысле кучи могут быть не равны как отображения различных множеств ключей, однако с точки зрения операции чтения они будут совпадать.

**Теорема 13.** Для любых символьных куч $\sigma, \sigma_1, \dots, \sigma_n$, произвольных непересекающихся ограничений $g_1, \dots, g_n$ и произвольной локации $x$ справедливо следующее:

$$union\langle g_1 \vee \dots \vee g_n, read(merge\langle g_i, \sigma_i\rangle \circ \sigma, x)\rangle = read(merge\langle g_i, \sigma_i \circ \sigma\rangle, x).$$

При помощи операции объединения возможно описать состояние программы на лист. 2 следующим образом:

$$\sigma_- = \{x \mapsto -LI(x)\}$$

$$\sigma = merge(\langle LI(x) \geq 0, \epsilon\rangle, \langle \neg(LI(x) \geq 0), \sigma_-\rangle)$$

$$read(\sigma, x) \overset{\text{Теор.11}}{=} union(\langle LI(x) \geq 0, read(\epsilon, x)\rangle, \langle \neg(LI(x) \geq 0), read(\sigma_-, x)\rangle) =$$
$$= union(\langle LI(x) \geq 0, LI(x)\rangle, \langle \neg(LI(x) \geq 0), -LI(x)\rangle).$$

```
Abs:
goto {x >= 0 -> Exit}
x := -x
Exit: halt
```

*Листинг 2. Объединение состояний.*

*Listing 2. Merge states.*

## 4.5 Запись в символьную кучу

До сих пор были рассмотрены операции с кучами как с готовыми объектами. Для того, чтобы строить кучу из пустого состояния $\epsilon$, необходима операция *записи* в символьную память. Для её определения воспользуемся следующим сокращением: $ite(c, a, b) \stackrel{\text{def}}{=} union(\langle c, a\rangle, \langle \neg c, b\rangle)$.

**Определение 8.** *Запись символьного значения $v$ в символьную локацию $y$ символьной кучи $\sigma$ — это символьная куча $write(\sigma, y, v)$, такая что для всех $x \in dom(write(\sigma, y, \cdot)) = dom(\sigma) \cup \{y\}$,*

$$(write(\sigma, y, v))(x) \stackrel{\text{def}}{=} ite(x = y, v, \sigma(x)).$$

Заметим, что инвариант кучи (1) для записей выполняется тривиально. Следующие теоремы показывают, что операция записи сохраняет свойство композициональности относительно других операций.

**Теорема 14.** Для любой символьной кучи $\sigma$, произвольных символьных локаций $x$, $y$ и любого символьного выражения $v$ справедливо следующее:

$$read(write(\sigma, y, v), x) = ite(x = y, v, read(\sigma, x)).$$

**Теорема 15.** Для любых символьных куч $\sigma$, $\sigma'$, произвольной символьной локации $y$ и произвольного символьного выражения $v$ справедливо следующее:

$$\sigma \circ write(\sigma', y, v) = write(\sigma \circ \sigma', \sigma \bullet y, \sigma \bullet v).$$

**Теорема 16.** Для любых символьных куч $\sigma_1, \ldots, \sigma_n$, любых непересекающихся ограничений $g_1, \ldots, g_n$, и произвольной символьной локации $y$ и символьного выражения $v$ справедливо следующее:

$$write(merge\langle g_i, \sigma_i\rangle, y, v) = merge\langle g_i, write(\sigma_i, y, v)\rangle.$$

## *5. Исчисление символьных куч*

Представленные операторы КСП уже позволяют описывать символьные состояния произвольных фрагментов кода без циклов в графе потока управления. Однако идея подстановки из контекстной кучи позволяет нам пойти дальше и определить исчисление, описывающее исполнение произвольных императивных программ с динамической памятью.

## 5.1 Обобщённые символьные кучи

Для начала определим формальный язык $Heap$ нашего исчисления. Термы языка $Heap$ будем называть *обобщёнными кучами*. Напомним, что $\Sigma$ — это множество всех символьных куч, которые были определены в предыдущем разделе.

Обобщённая куча может быть либо обычной символьной кучей (из $\Sigma$), которую мы будем называть *определённой*, либо объединением обобщённых куч по непересекающимся ограничениям, либо композицией обобщённых куч, либо записью в обобщённую кучу, либо неподвижной точкой цикла в графе потока управления, которую мы будем называть *рекурсивным состоянием*.

$$
\begin{aligned}
Heap ::= \ & \Sigma \\
& \mid Heap \circ Heap \\
& \mid merge(\langle guard, Heap\rangle^*) \\
& \mid write(Heap, loc, term) \\
& \mid Rec(id)
\end{aligned}
$$

*Рис. 7. Обобщённые кучи.*

*Fig. 7. Generalized heaps.*

Интуитивно, рекурсивные состояния программы — это состояния, зависящие от самих себя, т.е. чтение из которых требует предыдущую версию того же состояния. В общем случае такое состояние нельзя выразить в виде конечной композиции других состояний, поэтому для них вводится новый символ. Идентификатор $id$ в $Rec(id)$ уникальным образом описывает цикл в графе потока управления: он состоит из метки из исходного кода и некоторого.

Чтобы адаптировать операции КСП, необходимо расширить синтаксис символьных выражений, как показано на рис. 8.

$$
\begin{aligned}
term ::= \ & arith \mid loc \\
arith ::= \ & \mathbb{N} \mid arith \pm arith \mid -arith \mid LI^{arith}(\boldsymbol{Heap}, loc) \\
& \mid union^{arith}(\langle guard, arith\rangle^*) \\
loc ::= \ & null \mid 0x[0-9]^+ \mid ident \\
& \mid loc.FieldName \mid LI^{loc}(\boldsymbol{Heap}, loc) \\
& \mid union^{loc}(\langle guard, loc\rangle^*) \\
guard ::= \ & \top \mid \bot \mid \neg guard \mid guard \wedge guard \mid guard \vee guard \\
& \mid arith = arith \mid arith < arith \mid loc = loc
\end{aligned}
$$

*Рис. 8. Грамматика обобщённых символьных выражений.*

*Fig. 8. Symbolic generalized terms.*

На чтение из *определённой* кучи $read(\sigma, x) = LI(x)$ можно смотреть как на следующее сообщение: «в $\sigma$ недостаточно информации, чтобы узнать $x$, — требуется дополнительный контекст». Сама операция чтения построена таким образом, что мы можем однозначно сказать, было ли успешно чтение $x$ из $\sigma$, — и если было, то предъявить результат. При чтении из *обобщённой* кучи, например, $read(Rec(F), x)$, могут возникнуть сложности. Если циклический фрагмент кода по метке $F$ меняет содержимое $x$, то мы не можем его корректно прочитать — для этого необходимо заранее знать, сколько раз исполнится $F$, что в общем случае невозможно. Очевидно также, что мы не можем отбросить $Rec(F)$ и вернуть просто $LI(x)$.

Таким образом, основная идея расширения термов состоит в «запоминании» источника каждого символьного значения (выделено жирным на рис. 8). Такое расширение является корректным, поскольку старые символьные значения $LI(x)$ теперь станут $LI(\epsilon, x)$. Уточнение также может быть расширено: $\tau \bullet$

$LI(\sigma, x) = read(\tau \circ \sigma, \tau \bullet x)$. Это не нарушит свойства КСП, так как по теор. 3, мы имеем $\tau \bullet read(\sigma, x) = read(\tau \circ \sigma, \tau \bullet x)$.

## 5.2 Правила редукции

Правила редукции символьных куч представлены на рис. 9. Буквами $H$ обозначены элементы языка $Heap$, буквами $t$ — символьные термы. $H[A/X]$ означает одновременную подстановку $A$ вместо $X$ в обобщённую кучу $H$; здесь $A$ и $X$ могут быть как обобщёнными кучами, так и символьными термами.

$$\frac{t_1 = t_2}{H \to H[t_2/t_1]} \ (7)$$

$$H \to H[Body(id)/Rec(id)] \ (8)$$

$$\epsilon \circ H \to H$$

$$H \circ \epsilon \to H$$

$$H_1 \circ (H_2 \circ H_3) \to (H_1 \circ H_2) \circ H_3$$

$$H \circ merge\langle g_i, H_i \rangle \to merge\langle H \bullet g_i, H \circ H_i \rangle$$

$$merge\langle g_i, H_i \rangle \circ H \to merge\langle g_i, H_i \circ H \rangle$$

$$H \circ write(H', x, v) \to write(H \circ H', H \bullet x, H \bullet v)$$

$$write(merge\langle g_i, H_i \rangle, x, v) \to merge\langle g_i, write(H_i, x, v) \rangle$$

$$\frac{g \text{ невыполнимо}}{merge(\langle g, H \rangle \cup X) \to merge(X)}$$

$$\frac{g \text{ общезначимо}}{merge\langle g, H \rangle \to H}$$

*Рис. 9. Правила редукции.*

*Fig. 9. Reduction rules.*

Корректность всех правил, кроме (8), обоснована в разд. 4. Правила (7) и (8) представляют собой аналог $\alpha$-конверсии и $\beta$-редукции в $\lambda$-исчислении. $Body(x)$ представляет собой описание поведения участка кода, которому соответствует обобщённая куча $Rec(x)$. Если посмотреть на исчисление символьных куч как на некоторый язык программирования, то $Rec(x)$ соответствовало бы имени функции, а $Body(x)$ — её телу. Обобщённую кучу назовём *нередуцируемой*, если к ней нельзя применить ни одного правила за исключением (7). $H \to^n H'$ означает, что куча $H$ редуцируется в $H'$ за $n$ шагов. $H \to^* H'$ означает, что существует $n \geq 0$, что $H \to^n H'$.

Теперь продемонстрируем, как можно представлять символьные состояния циклических фрагментов кода на примере с лист. 3. Так как выход из цикла зависит от получаемого списка $p$, и его длина заранее неизвестна, то состояние этого фрагмента нельзя представить в виде какой-либо конечной композиции определённых символьных состояний. Однако можно построить обобщённую кучу $Body(Inc)$, описывающую поведение метки $Inc$.

```
1. Inc:
2. goto {p = null -> Exit}
3. p.Key := p.Key + 1
4. p := p.Next
5. goto {true -> Inc}

6. Exit: halt
```

*Листинг 3. Фрагмент кода с циклом в графе потока управления.*

*Listing 3. Code snippet with a cycle in a control flow graph.*

Пусть $\sigma_0$ — некоторая символьная куча, представляющая начальное состояние исполнения, а обобщённая куча $Rec(Inc)$ соответствует метке $Inc$. Тогда поведение всего кода в лист. 3 на состоянии $\sigma_0$ будет описываться обобщённой кучей $\sigma_0 \circ Rec(Inc)$. Применение правил редукции к $\sigma_0 \circ Rec(Inc)$ будет соответствовать вычислению кода с метки $Inc$ на состоянии $\sigma_0$. Покажем это на примере.

Пусть $\sigma_0 = \{p \mapsto 0x1, 0x1.Key \mapsto 10, 0x1.Next \mapsto 0x2, 0x2.Key \mapsto 20, 0x2.Next \mapsto null\}$.

Опишем теперь поведение циклического региона, помеченного $Inc$. В начале исполнения происходит ветвление по условию $p = null$. Если $p \neq null$, то стр. 3-5 увеличивают значение ключа в узле связного списка и переходят к следующему элементу. Поведение этого участка кода можно описать символьным объединением двух эффектов: пустого эффекта $\epsilon$ (т.к. переход на стр. 2 не меняет состояния) и эффекта $\sigma$ нерекурсивного кода на стр. 3-4, где $\sigma = \{LI(p).Key \mapsto LI(LI(p).Key) + 1, p \mapsto LI(LI(p).Next)\}$.

Таким образом, поведение региона $Inc$ описывается обобщённой кучей

$$Body(Inc) = merge(\langle LI(p) = null, \epsilon\rangle, \langle LI(p) \neq null, \sigma \circ Rec(Inc)\rangle).$$

Теперь опишем процесс редукции кучи $\sigma_0 \circ Rec(Inc)$.

$$\sigma_0 \circ Rec(Inc) \to \sigma_0 \circ Body(Inc) \to$$
$$merge(\langle \sigma_0 \bullet (LI(p) = null), \sigma_0 \circ \epsilon\rangle, \langle \sigma_0 \bullet (LI(p) \neq null), \sigma_0 \circ (\sigma \circ Rec(Inc))\rangle) \to^4$$
$$merge(\langle 0x1 = null, \sigma_0\rangle, \langle 0x1 \neq null, (\sigma_0 \circ \sigma) \circ Rec(Inc)\rangle) \to^2 \sigma_1 \circ Rec(Inc) \to^2$$
$$merge(\langle \sigma_1 \bullet (LI(p) = null), \sigma_1 \circ \epsilon\rangle, \langle \sigma_1 \bullet (LI(p) \neq null), \sigma_1 \circ (\sigma \circ Rec(Inc))\rangle) \to^4$$
$$merge(\langle 0x2 = null, \sigma_1\rangle, \langle 0x2 \neq null, (\sigma_1 \circ \sigma) \circ Rec(Inc)\rangle) \to^2 \sigma_2 \circ Rec(Inc) \to^2$$
$$merge(\langle \sigma_2 \bullet (LI(p) = null), \sigma_2 \circ \epsilon\rangle, \langle \sigma_2 \bullet (LI(p) \neq null), \sigma_2 \circ (\sigma \circ Rec(Inc))\rangle) \to^4$$
$$merge(\langle null = null, \sigma_2\rangle, \langle null \neq null, (\sigma_2 \circ \sigma) \circ Rec(Inc)\rangle) \to^2 \sigma_2$$

Здесь

$$\sigma_1 \stackrel{\text{def}}{=} \sigma_0 \circ \sigma = \{p \mapsto 0x2, 0x1.Key \mapsto 11, 0x1.Next \mapsto 0x2,$$
$$0x2.Key \mapsto 20, 0x2.Next \mapsto null\},$$
$$\sigma_2 \stackrel{\text{def}}{=} \sigma_1 \circ \sigma = \{p \mapsto null, 0x1.Key \mapsto 11, 0x1.Next \mapsto 0x2,$$
$$0x2.Key \mapsto 21, 0x2.Next \mapsto null\}.$$

Обобщённая куча $\sigma_2$ не редуцируема (заметим, что нередуцируемым будет любой элемент $\Sigma$). $\sigma_0$ и $\sigma_1$ представляют собой состояния изначальной императивной программы в процессе её исполнения, а $\sigma_2$ — её конечное состояние (при запуске на $\sigma_0$).

Исчисление символьных куч позволяет описывать произвольные поведения программ с динамической памятью без потери информации.

## *6. Композициональное символьное исполнение*

В данном разделе предложен алгоритм композиционального символьного исполнения, который позволяет автоматически проверять достижимость ошибок при произвольном графе потока управления. Он основан на подходе символьного исполнения программ [22].

### 6.1 Метод описания путей в графе потока управления

Этот метод является основным в предлагаемом алгоритме, т.к. выполняет описание *всех* путей в графе потока управления при помощи введения минимального числа рекурсивных состояний. Данный метод, в частности, позволяет *автоматически* строить обобщённые символьные кучи $Body(id)$, описывающие поведения циклических фрагментов программы.

Прямолинейным способом построения обобщённых куч по императивной программе является введение куч $Rec(l)$ для каждой инструкции $l$ и взятие композиции со всеми $Rec(l')$, в которые есть переход из инструкции $l$. Однако такой подход порождает слишком большую систему взаимно-рекурсивных определений: фактически, количество символов-абстракций $Rec(l)$ было бы равно количеству инструкций в программе. В данном разделе описывается метод описания всех возможных путей исполнения программы через введение меньшего числа символов-абстракций.

**Определение 9.** *Вершинами* $V_G$ графа потока управления $G$ будут номера $l$ инструкций, а *рёбрами* $E_G$ — пары номеров $(l_x, l_y)$, указывающие на возможность передачи управления от инструкции $l_x$ к инструкции $l_y$. В графе потока управления существует *начальная* вершина, которая соответствует первой инструкции программы и в которую не ведёт ни одно ребро. Согласно грамматике демо-языка (см. рис. 1), большинство рёбер будут иметь вид $(l, succ(l))$. Однако благодаря оператору «**goto**» возможны переходы к произвольным инструкциям, кроме начальной.

**Определение 10.** Проведём обход графа $G$ в глубину и для каждой вершины $v$ вычислим время «выхода» $time(v)$ из обхода. Вершина $l$ называется *рекурсивной*, если существует ребро $(l', l)$ такое, что $time(l') \geq time(l)$. Множество рекурсивных вершин будем обозначать $RV$.

Для описания путей необходимо ввести операцию *конкатенации* двух путей в графе. Неформально, конкатенация путей $p_1$ и $p_2$ — это путь $p_1 \circ p_2$, который содержит рёбра пути $p_1$, за которыми следуют ребра пути $p_2$. Конкатенация двух множеств путей $P_1$ и $P_2$ определяется через конкатенацию двух путей:

$P_1 \circ P_2 = \{p_1 \circ p_2 | p_1 \in P_1, p_2 \in P_2\}$. Символ $\varepsilon$ означает *пустой путь*, а символ $\cup$ — объединение множеств путей.

Предлагаемый метод позволяет описывать в точности все пути в произвольном графе потоке управления при помощи рекурсивных символов и их рекурсивных описаний, на базе которых далее будут строиться обобщённые символьные кучи.

$$\Pi(u,v,D) = \bigcup_{(u,v)\in E_G} \{(u,v)\} \cup \bigcup_{\substack{t\notin RV\cup\{v\} \\ (u,t)\in E_G}} (u,t)\circ \Pi(t,v,D) \cup \bigcup_{\substack{t\in RV\setminus(D\cup\{v\}) \\ (u,t)\in E_G}} (u,t)\circ Rec(t, D\cup\{t\})\circ \Pi(t,v,D\cup\{t\})$$

$$Rec(u,D) = \{\varepsilon\} \cup \Pi(u,u,D)\circ Rec(u,D)$$

Символом $\Pi(u,v,D)$ обозначим множество путей из вершины $u$ в вершину $v$, параметризованное множеством *пройденных* рекурсивных вершин $D$. Множество $D$ ограничивает переходы по рёбрам: ребро $(l_x, l_y)$ является «допустимым», если оно ведёт в конечную вершину (т.е. $l_y = v$) или $l_y \neq v$ и вершина $l_y$ не была ещё посещена (т.е. $l_y \notin D$). *Рекурсивным символом* $Rec(u,D)$ обозначим множество путей-циклов из вершины $u$ в вершину $u$ с множеством $D$, имеющим тот же смысл, как и для $\Pi(u,v,D)$.

Интуитивно, $\Pi(u,v,D)$ соответствует символьным кучам, полученным в результате символьного исполнения программы от инструкции с номером $u$ до инструкции с номером $v$, когда исполнение не посещало инструкции с номерами из множества $D\setminus\{v\}$. В свою очередь, *рекурсивный символ* $Rec(u,D)$ соответствует *обобщённой символьной куче* $Rec(id)$, у которой уникальным идентификатором $id$ является пара $(u,D)$, а его описание — это обобщённая символьная куча $Body(u,D)$. Кроме того, оператор $\circ$ соответствует операции *композиции* состояний, символ $\cup$ — операции *объединения* состояний ($merge$), а $\varepsilon$ — пустой куче $\epsilon$.

Покажем на примере, как можно описать все пути в графе потока управления при помощи итеративного построения $\Pi$ и $Rec$. На рисунке ниже представлен пример графа потока управления программы с вложенными циклами. Для простоты изложения номер вершины $v$ равен времени $time(v)$. На рисунке вершины 1 и 2 являются рекурсивными. Ниже представлено описание всех путей в графе из 0 в 5, использующее рекурсивные символы.

$$\Pi(0,5,\varnothing) = (0,1)\circ Rec(1,\{1\})\circ(1,2)\circ Rec(2,\{1,2\})\circ(2,3)\circ(3,4)\circ(4,5)$$

$$Rec(1,\{1\}) = (1,2)\circ Rec(2,\{1,2\})\circ(2,3)\circ(3,4)\circ(4,1)\circ Rec(1,\{1\})\cup\{\varepsilon\}$$

$$Rec(2,\{1,2\}) = (2,3)\circ(3,2)\circ Rec(2,\{1,2\})\cup\{\varepsilon\}$$

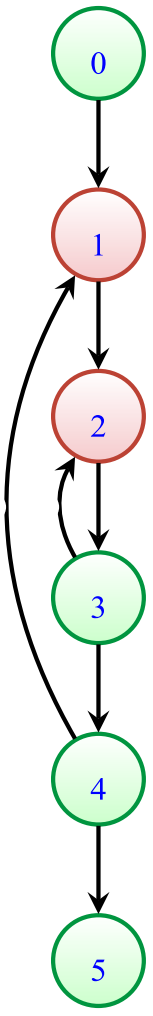

$\Pi(0,5,\varnothing)$ обозначает все пути из вершины 0 в вершину 5. При переходе по ребру $(0,1)$ метод попадает в вершину 1. Так как она *рекурсивная* (поскольку лежит в $RV$), метод вводит для неё символ $Rec(1,\{1\})$ и начинает создавать *рекурсивное* описание этого символа.

Это описание начинается с перехода в вершину 2 и введения рекурсивного символа для неё. Поскольку были пройдены обе *рекурсивные* вершины, то при создании символа $Rec(2,\{1,2\})$ множество $D = \{1,2\}$. Рекурсивное определение для $Rec(2,\{1,2\})$ выглядит следующим образом: все пути из 2 в 2, не проходящие через $D = \{1,2\}$ в середине пути, — это повторения пути $(2,3) \circ (3,2)$. Кроме того, любое множество $Rec(\cdot)$ путей из себя в себя содержит пустой путь $\varepsilon$.

Далее продолжается описание $Rec(1,\{1\})$: после перехода из 1 в 2 происходит конкатенация пути $(1,2)$ и путей $Rec(2,\{1,2\})$ из 2 в себя, а к множеству $D$ добавляется вершина 2. После этого путь, возвращающийся в вершину 1, очевиден: $(2,3) \circ (3,4) \circ (4,1)$.

Затем процесс возвращается к построению $\Pi(0,5,\varnothing)$. После перехода по ребру $(0,1)$ и создания символа $Rec(1,\{1\})$ к множеству D добавляется вершина 1. Затем происходит переход по ребру $(1,2)$ и добавление соответствующего символа $Rec(2,\{1,2\})$, а также к множеству $D$ добавляется 2. Поскольку описание для символа $Rec(2,\{1,2\})$ уже было построено при построении описания $Rec(1,\{1\})$, то метод не будет строить его заново. Далее следует тривиальный путь до вершины 5: $(2,3) \circ (3,4) \circ (4,5)$.

Таким образом, метод позволяет описывать все пути в графе потока управления и только их. Формальное доказательство этого факта приведено в прил. C.

## 6.2 Алгоритм композиционального символьного исполнения

Благодаря соответствию между рекурсивными символами и их описаниями, с одной стороны, и обобщёнными кучами $Rec(id)$ и $Body(id)$, с другой стороны, можно получить *алгоритм автоматической проверки достижимости ошибок в программах с динамической памятью и произвольными графами потока управления*, не раскручивающий отношение перехода.

Предлагаемый алгоритм использует символьное исполнение и операции над символьными кучами (см. лист. 4). Также он использует оракул $SAT$, проверяющий достижимость веток исполнения. Для данного алгоритма важно понятие *состояния исполнения*.

**Определение 10.** *Состояние исполнения* — это кортеж $(l, pc, \sigma, D)$, где $l$ — номер инструкции, $pc$ — *условие пути* (path condition — символьная формула, описывающая ограничения на достижимость инструкции), $\sigma$ — символьная куча, $D$ — множество посещённых рекурсивных вершин.

Важнейшей функцией алгоритма является $Exec$, которая может быть вызвана либо из *начальной* (стр. 2), либо из *рекурсивной* вершины (стр. 51). В первом случае её результатом является символьная куча, соответствующая путям исполнения, которые привели к инструкциям $\boldsymbol{halt}$ (стр. 10), а также выводится множество путей, приводящих к ошибкам (стр. 53). Во втором случае результатом является символьная куча $Body(l_0, D_0)$, описывающая поведения циклического участка графа потока управления. В стр. 37-40 происходит добавление *обобщённого состояния*, представляющего собой композицию *построенного состояния* $\sigma'$ *с рекурсивным состоянием* $Rec(l_0, D_0)$. В конце функции (стр. 52) добавляется завершающее состояние для случая, когда исполнение не вернулось в *рекурсивную* вершину $l_0$, соответствующее *пустому пути* $\varepsilon$ из определения рекурсивных символов в методе описания путей в графе.

Алгоритм выбирает следующее состояние исполнения из рабочего множества ($pickNext$), затем исполняет соответствующую инструкцию, порождая новые состояния исполнения (стр. 9-35), и добавляет их в рабочее множество, объединяя те состояния исполнения, у которых равны номера инструкций $l$ и совпадают множества *посещённых* рекурсивных вершин $D$ (стр. 47-50). Стоит отметить, что предлагаемый алгоритм не раскручивает циклы, а вводит рекурсивные состояния $Rec(l_0, D_0)$ (стр. 12) и обобщённые кучи $Body(l_0, D_0)$ (стр. 46) для их описания. Все обобщённые кучи $Body(\cdot,\cdot)$ используются для определения выполнимости ограничений пути (стр. 21, 23, 31, 34).

Функция $Eval(\sigma, expression)$ вычисляет выражения, трактуя арифметические и булевы операции стандартным образом и читая переменные из состояния $\sigma$.

Стоит заметить, что в качестве побочного эффекта $Eval(\sigma, expression)$ может добавить к множеству $Errors$ новое условие пути, защищающее обращение к нулевому адресу. Для вычисления $Eval(\sigma, \boldsymbol{new}\,\{Field_i \to Expr_i\})$ будет создан новый уникальный адрес $0xNNN$[3], все $Expr_i$ вычислятся в $v_i$ и состояние будет итеративно обновлено: $\sigma' = write(\sigma, 0xNNN.Field_i, v_i)$. Функция $Eval$ вернёт новое состояние $\sigma'$ и адрес созданного объекта $0xNNN$. Например, для фрагмента на лист. 5 может быть получено следующее новое состояние:

$$\begin{array}{lll} 0x40.K \mapsto 30 & 0x41.K \mapsto 10 & 0x42.K \mapsto 50 \\ 0x40.L \mapsto 0x41 & 0x41.L \mapsto null & 0x42.L \mapsto null \\ 0x40.R \mapsto 0x42 & 0x41.R \mapsto null & 0x42.R \mapsto null \end{array}$$

[3] В текущем изложении некорректно обрабатывается случай с выделением объекта в циклическом регионе; для корректной работы определение уточнения должно быть изменено: $\sigma \bullet 0xNNN$ должно порождать *новый* адрес $0xMMM$

```
∀l ∈ RV, ∀D  Body(l, D) ← ϵ;
return Exec(start, ∅);
Function Exec(l_O : Vertex, D_O : Vertex set)
    pc_r, σ_r ← (⊥, ϵ);
    W ← {(l_O, ⊤, ϵ, D_O)}; Errors ← ∅;
    while W ≠ ∅ do
        (l, pc, σ, D), W ← pickNext(W);
        S ← ∅;
        switch instr(l)
            case halt:
                if l_O ∉ RV then
                    pc_r ← pc_r ∨ pc;
                    σ_r ← merge((pc_r , σ_r ), (pc,σ));
            case fail:   Errors ← Errors ∪ {pc};
            case ident := expression
                σ, value ← Eval(σ, expression);
                S ← {(succ(l), pc, write(σ, ident, value))};
            case Location.field := expression
                σ, value ← Eval(σ, expression);
                σ, loc ← Eval(σ, Location);
                if SAT(Body, σ, pc ∧ loc ≠ null) then
                    S ← {(succ(l), pc ∧ loc ≠ null, write(σ, loc.field, value))};
                if SAT(Body, σ, pc ∧ loc = null) then
                    Errors ← Errors ∪ {pc ∧ loc = null};
            case label : statement
                S ← {(succ(l), pc, σ)};
            case goto labels
                guard_succ ← ⊤;
                forall (expression → l') ∈ labels do
                    σ, guard ← Eval(σ, expression);
                    if SAT(Body, σ, pc ∧ guard ∧ guard_succ) then
                        S ← S ∪ {(l', pc ∧ guard ∧ guard_succ, σ)};
                    guard_succ ← guard_succ ∧ ¬guard;
                if SAT(Body, σ, pc ∧ guard_succ) then
                    S ← S ∪ {(succ(l), pc ∧ guard_succ, σ)};
        forall (l', pc', σ') ∈ S do
            if l' = l_O then
                σ' ← σ' ∘ Rec(l_O, D_O);
                pc_r, σ_r ← (pc_r ∨ pc', merge ((pc_r, σ_r), (pc', σ')));
                continue;
            elif l' ∈ D then continue;
            elif l' ∈ RV then
                D ← D ∪ {l'};
                σ' ← σ' ∘ Rec(l', D);
                if Body(l', D) = (⊥, ϵ) then
                    Body(l', D) ← Exec(l', D);
            if ∃(l'', pc'', σ'', D'') ∈ W : l' = l'' ∧ D = D'' then
                W ← W \ {(l'', pc'', σ'', D'')};
                W ← W ∪ {(l', pc' ∨ pc'', merge( (pc', σ'), (pc'', σ'')) , D)};
            else W ← W ∪ {(l', pc', σ', D)};
    if l_O ∈ RV then
        return merge (pc_r, σ_r), (¬pc_r, ϵ) ;
    print Errors;
    return σ_r
```

*Листинг 4. Алгоритм композиционального символьного исполнения.*

*Listing 4. Compositional symbolic execution algorithm.*

```
1.  x = new {K = 30;
2.           L = new {K = 10; L = null; R = null};
3.           R = new {K = 50; L = null; R = null}}
```

*Листинг 5. Программа, выделяющая память.*

*Listing 5. Heap-allocating program.*

## 6.3 Корректность алгоритма композиционального символьного исполнения

**Определение 12.** *Замкнутым* назовём терм, не содержащий $LI(\cdot)$.

*Конкретная куча* — это (тотальное) отображение из замкнутых локаций в замкнутые термы. Множество конкретных куч обозначим за $\Sigma_G$.

Конкретная куча представляет собой состояние динамической памяти при конкретном исполнении программы. Заметим, что $(\Sigma_G,\circ)$ — правый идеал в моноиде $(\Sigma,\circ)$: если $\sigma \in \Sigma_G, \tau \in \Sigma$, то $\sigma \circ \tau \in \Sigma_G$. Этот факт позволяет легко доказать следующее утверждение.

**Утверждение 2.** Если для $\sigma \in \Sigma_G$ и обобщённой кучи $H$, $\sigma \circ H \to^* H'$ для некоторой нередуцируемой $H' \in Heap$, то $H' \in \Sigma_G$.

Пусть $T: \mathbb{N} \times \Sigma_G \to \mathbb{N} \times \Sigma_G$ — отношение перехода некоторой программы на демо-языке (т.е. отображение, которое номеру инструкции и состоянию программы сопоставляет следующую инструкцию и состояние, полученное исполнением входной инструкции на входном состоянии). Обозначим $T^n(l,\sigma) \stackrel{\text{def}}{=} \underbrace{T(\dots T}_{n \text{раз}}(l,\sigma))$.

Следующая теорема (которую мы оставляем без доказательства) говорит о корректности алгоритма на лист. 4.

**Теорема 17.** Пусть $T$ — отношение перехода программы $P$ на демо-язык е, $l_0$ — номер начальной инструкции, $F$ — множество номеров инструкций ***halt*** и ***fail*** в программе, $\sigma_0 \in \Sigma_G$, $H \stackrel{\text{def}}{=} Exec(l_0, \varnothing)$. Также допустим, что оракул $SAT$ всегда отвечает правильно. Тогда $T^n(l_0,\sigma_0) = (f,\tau)$ для некоторых $n \in \mathbb{N}, f \in F, \tau \in \Sigma_G$ т. и т. т., к. $\sigma_0 \circ H \to^* \tau$.

## *7. Трансляция символьных куч в чистые функции*

Для проверки достижимости некоторого пути исполнения программы, алгоритм из разд. 6 обращается к функции-оракулу $SAT$. В данном разделе мы определяем $SAT(Body,\sigma,g)$ и тем самым завершаем построение композициональной процедуры верификации. Мы сведём задачу выполнимости ограничения $g$ к задаче доказательства безопасности функциональной программы без эффектов, состоящей из чистых функций второго порядка[4].

## 7.1 Оператор Find

Этот оператор, определённый в разд. 4.3, играет важную роль в трансляции обобщённых куч в чистые функции, обобщая чтение и композицию символьных куч. Сформулируем его основное свойство.

**Утверждение 3.** Для всех определённых символьных куч $\sigma$, $\sigma'$, $\tau$ таких, что для каждого символьного выражения $e$, выполняется $(\tau \circ \sigma) \bullet e = \tau \bullet (\sigma \bullet e)$, и всех символьных выражений $x \in loc$, $d \in term$, верно следующее:

$$\tau \bullet find(\sigma', x, \sigma, d) = find(\sigma', \tau \bullet x, \tau \circ \sigma, \tau \bullet d).$$

Далее, можно заметить, что:

$$read(\sigma, x) = find(\sigma, x, \epsilon, LI(x)) = find(\sigma, x, \epsilon, read(\epsilon, x)),$$
$$(\sigma \circ \sigma')(x) = find(\sigma', x, \sigma, \sigma(x)) = find(\sigma', x, \sigma, read(\sigma, x)).$$

Это даёт возможность определить оператор $find$ для обобщённых куч следующим образом (обозначив прежний $find$ как $find^{\Sigma}$):

$$find(\sigma, x, \tau) \overset{\text{def}}{=} \begin{cases} find^{\Sigma}(\sigma, x, \tau, find(\tau, x, \epsilon)), & \text{если } \sigma \in \Sigma \\ LI(\tau \circ \sigma, x), & \text{иначе} \end{cases} \quad (10)$$

## 7.2 Трансляция обобщённых куч в функции второго порядка

Будем говорить, что символьный терм $t$ находится в *нормальной форме*, если он содержит объединения только на верхнем уровне, т.е. $t = union(\langle g_1, t_1\rangle, \dots, \langle g_n, t_n\rangle)$ и ни одно из ограничений $g_i$ и ни один из термов $t_i$ не содержат внутри $union$. Будем также говорить, что *ограничение* находится в нормальной форме, если оно не содержит объединений. Каждое символьное выражение может быть нормализовано: по утв. 1 и (6), вложенные объединения могут быть линеаризованы. Если $t$ не содержит объединений, тогда его нормальной формой будем называть $union\langle \top, t\rangle$. По определению, ограничение $g \equiv union(\langle g_1, c_1\rangle, \dots, \langle g_n, c_n\rangle)$ может быть переписано в $(g_1 \wedge c_1) \vee \dots \vee (g_n \wedge c_n)$.

Рассмотрим символьную ячейку $LI(\sigma, x)$. Заметим, что такие ячейки с $\sigma \neq \epsilon$ появляются только в последней ветке определения (10), т.е. можно рассматривать $LI(\sigma, x)$ как $find(\sigma, x, \epsilon)$. Уточнение такого выражения в контексте $\tau$ даст $\tau \bullet find(\sigma, x, \epsilon) \overset{\text{утв.3}}{=} find(\sigma, \tau \bullet x, \tau)$. Это даёт возможность транслировать символьные выражения в функции второго порядка.

Далее, с помощью $\tau$ обозначим функцию первого порядка «чтение из контекстной кучи». Преобразование символьного выражения $e$ в выражение функционального языка при контекстной куче $\tau$ обозначим как $[\![e]\!]_\tau$. Это преобразование состоит из трёх следующих шагов.

[4] Функцией *первого порядка* называется функция, которая не принимает в аргументы другие функции. Функцией *второго порядка* называется функция, которая принимает в аргументы функции только первого порядка — и не выше.

1. Нормализация $e$ и преобразование верхнеуровневого объединения в конструкцию ветвления.
2. Замена всех ячеек $LI(\sigma, x)$ на $find(\sigma, [\![x]\!]_\tau, \tau)$.
3. Специализация оператора $find$ согласно правилам (10). На этом шаге все термы вида $find(\sigma, x, \tau)$ транслируются в применения функций второго порядка $\text{find}_\sigma$. Телом функции $\text{find}_\sigma$ будет результат применения этих трёх шагов к соответствующему правилу (10). При появлении композиции $\sigma \circ \sigma'$ контекстное состояние становится частичным применением $\text{find}_\sigma$ к текущему контекстному состоянию $\tau$.

Вместо формального описания целевого функционального языка программирования и алгоритма трансляции мы продемонстрируем процесс трансляции на примере. Допустим, необходимо ответить на запрос $SAT(Body, Rec(f), LI(\epsilon, a) * 3 < 17)$. Пусть $Body(f) = merge(\langle c, \epsilon\rangle, \langle \neg c, \sigma \circ Rec(f)\rangle)$, где $\sigma$ — это некоторая обобщённая куча. Тогда необходимо проверить выполнимость ограничения $g = (Rec(f) \bullet (LI(\epsilon, a) * 3) < 17) = (LI(Rec(f), a) * 3 < 17)$.

Сначала определим контекстную функцию первого порядка $\tau_0$, которая будет принимать адрес и выполнять ленивое инстанцирование символьных локаций, т.е. возвращать недетерминированные значения. Далее, вычислим $[\![g]\!]_{\tau_0}$.

Первый шаг не порождает условных конструкций. После второго шага выражение $g$ становится следующим: $find(Rec(f), a, \tau_0) * 3 < 17$. Третий шаг порождает новую функцию второго порядка $\text{find}_{Rec(f)}$. Таким образом, закодированное значение $g$ будет: $[\![g]\!]_{\tau_0} = (find_{Rec(f)}\ \tau_0\ a) * 3 < 17$.

Проверить выполнимость $g$ — это то же самое, что проверить безопасность программы “assert(¬ g)”.

Теперь мы должны задать тела полученных функций $find$. Пусть $\text{find}_f$ принимает контекстную функцию первого порядка $\tau$ и локацию $x$. Тело функции $\text{find}_{Rec(f)}$ мы получим, применяя шаги 1-3 к $Body(f)$:

$$find(merge(\langle c, \epsilon\rangle, \langle \neg c, \sigma \circ Rec(f)\rangle), x, \tau) \overset{(10)}{=}$$
$$= ite(\tau \bullet c, find(\epsilon, x, \tau), find(\sigma \circ Rec(f), x, \tau))$$

Это объединение будет нормализовано и транслируется в ветвление в теле $\text{find}_{Rec(f)}$; ленивые ячейки в $c$ заменятся применениями $find$, которые будут также специализированы. Итеративное применение этих шагов даст код для $g$, представленный ниже.

Идея такой трансляции заключается в том, что операции композиции могут быть заменены частичными применениями функций. Это позволяет сохранять справедливость того факта, что контекстные функции не поднимаются выше первого порядка. Таким образом получается трансляция в чистые функции второго порядка.

```
assert(not((find_Rec(f) τ a) * 3 < 17))
find_Rec(f) τ x =
  if ⟦g⟧_τ then find_ϵ τ x
  else find_σ∘Rec(f) τ x
find_ϵ τ x = τ x
find_σ∘Rec(f) τ x = find_Rec(f) (find_σ τ) x
find_σ τ x = …
```

**Замечание.** Есть несколько способов улучшить эту трансляцию. Во-первых, можно специализировать не только по куче, но и по типу локации, что даст более специализированные функции. Это также необходимо, чтобы получить из алгоритма трансляции типобезопасные функции. Во-вторых, полученная программа может быть частично исполнена, чтобы удалить тривиальные чтения, как, например, $\text{find}_{\epsilon}$. В-третьих, именованные локации могут передаваться как обычные аргументы, так как их адреса никогда не меняются. Эти три улучшения позволяют получить автоматическую трансляцию, чьи результаты будут схожи с приведённой в прил. А.

## 7.3 Корректность трансляции в чистые функции

Следующая теорема (которую мы оставляем без доказательства) говорит о корректности алгоритма из разд. 7.2.

**Теорема 18.** Пусть $H \in Heap$, $\sigma_0 \in \Sigma_G$, $g$ — символьное ограничение. Тогда $\tau \bullet g$ выполнимо для некоторого $\tau \in \Sigma_G$, такого что $\sigma_0 \circ H \to^* \tau$, т. и т. т., к. небезопасна функциональная программа, полученная из $\sigma_0 \circ H$ и $g$ применением алгоритма кодирования обобщённых куч из разд. 7.2.

Таким образом, сводя воедино корректность символьного исполнения и кодирования, получаем следующую теорему.

**Теорема 19.** Пусть $isSafe(p)$ — оракул, проверяющий безопасность функциональных программ, который всегда возвращает верный результат. Тогда программа на демонстрационном языке безопасна т. и т. т., к. алгоритм с лист. 4 выводит $Errors = \varnothing$.

*Доказательство.* Следует из теор. 17 и теор. 18.

## 7.4 Верификация функциональных программ без эффектов

Для доказательства безопасности функциональных программ без эффектов можно применить различные классические техники. Одной из самых успешных является *вывод уточнённых типов* (refinement type inference) [10, 23, 24, 25]. Фреймворки вывода уточнённых типов строят индуктивные инварианты функциональных программ высших порядков из инвариантов первого порядка над значениями с закрытыми типами (ground-types). Более точное описание этого процесса содержится в [25].

Тот факт, что получаемые в результате нашей трансляции функции не выше второго порядка, позволяет специализировать и оптимизировать процедуру вывода уточнённых типов. В контексте нашей работы, наиболее интересны *композициональные* фреймворки вывода уточнённых типов. Примером такого фреймворка является [24].

## 8. Заключение

В работе представлен подход к композициональному точному анализу программ с динамической памятью. Подход сводит задачу доказательства корректности таких программ к задаче вывода уточняющих типов функциональных программ через построение обобщённых куч, описывающих эффект программы на произвольном состоянии. Имея хорошую модель памяти [26], подход легко адаптировать к промышленным языкам программирования, и тем самым свести задачу формальной верификации программ на таких языках к решению рекурсивно-логических соотношений. Построение практичного верификатора языка C#, основанного на представленном подходе, будет описано в следующих работах. Модель композициональной символьной памяти может также служить хорошим подспорьем для языка спецификации свойств императивных программ с динамической памятью.

Вне контекста формальной верификации работа может рассматриваться как построение интересного соответствия между императивными программами с динамической памятью и чистыми функциями: описанное в статье сведение способно по императивной программе с динамической памятью породить эквивалентную программу на чистом функциональном языке, причём порождённые функциональные программы неочевидным образом кодируют операции над динамической памятью, а композициональное сведение позволяет порождать код функций, не зависящий от её отдельных вызовов.

Предложенный в статье подход выглядит перспективным при обеспечении качества встроенных систем реального времени [28], а также при разработке операционных систем реального времени [29]. Также перспективным является интеграция данного подхода со средствами визуального моделирования ПО, в частности, при верификации исполняемых визуальных спецификаций [27].

## Список литературы

# Automatic verification of heap-manipulating programs

[1] *Yu.O. Kostyukov <kostyukov.yurii@gmail.com>*
[1] *K.A. Batoev <konstantin.batoev@gmail.com>*
[1] *D.A. Mordvinov <dmitry.mordvinov@jetbrains.com>*
[1] *M.P. Kostitsyn <mishakosticyn@yandex.ru>*
[1] *A.V. Misonizhnik <misonijnik@gmail.com>*
[1] *Saint Petersburg State University,*
*7/9 Universitetskaya nab., St. Petersburg, 199034, Russia.*
[2] *JetBrains Research.*

**Abstract.** Theoretical foundations of compositional reasoning about heaps in imperative programming languages are investigated. We introduce a novel concept of compositional symbolic memory and its relevant properties. We utilize these formal foundations to build up a compositional algorithm that generates generalized heaps, terms of symbolic heap calculus, which characterize arbitrary cyclic code segments. All states inferred by this calculus precisely correspond to reachable states of the original program. We establish the correspondence between inference in this calculus and execution of pure second-order functional programs.

# A. Пример трансляции программы с динамической памятью в Haskell[5]

```c
1   #include <stdlib.h>
2   #include <stdio.h>
3
4   typedef struct node {
5     int key;
6     struct node *next;
7   } node_t;
8
9   void inc(node_t *node) {
10    if (node) {
11      node->key += 1;
12      inc(node->next);
13    }
14  }
15
16  void main() {
17    node_t  *a  =  malloc(sizeof(node_t));
18    node_t  *b  =  malloc(sizeof(node_t));
19    node_t  *c  =  malloc(sizeof(node_t));
20    a->key  =10;
21    b->key = 20;
22    c->key  =30;
23    a->next  = b;
24    b->next  = c;
25    c->next  = NULL;
26    inc(a);
27    printf("%d\n", c->key);
28  }
```

```haskell
data Field = Key | Next
type Addr = Int
type Loc = (Addr, Field)

ub _ = error "Undefined behaviour"

-- Heap before the recursive call of "inc"
h1_int :: (Loc -> Int) -> Addr -> Loc -> Int
h1_int ctx node loc@(addr, Key) | addr == node = ctx loc + 1
h1_int ctx _ loc = ctx loc

h1_ptr :: (Loc -> Addr) -> Loc -> Addr
h1_ptr ctx loc = ctx loc

-- The effect of "inc"
inc_int :: (Loc -> Int) -> (Loc -> Addr) -> Addr -> Loc -> Int
inc_int ctx_int ctx_ptr 0 loc = ctx_int loc
inc_int ctx_int ctx_ptr node loc =
    let next = ctx_ptr (node, Next) in
   inc_int (h1_int ctx_int node) (h1_ptr ctx_ptr) next loc

-- Heap before the call of "inc" from "main"
h2_int :: (Loc -> Int) -> Loc -> Int
h2_int _ (1, Key) = 10
h2_int _  (2, Key)  = 20
h2_int _ (3, Key) = 30
h2_int ctx loc = ctx loc

h2_ptr :: (Loc -> Addr) -> Loc -> Addr
h2_ptr _ (1, Next) = 2
h2_ptr _  (2, Next)  = 3
h2_ptr _ (3, Next) = 0
h2_ptr ctx loc = ctx loc

main =
   print $ show $ inc_int (h2_int ub) (h2_ptr ub) 1 (3, Key)
```

Аналог алгоритма композиционального символьного исполнения на лист. 4 породит кучу $h_2 \circ Rec(inc)$ для функции $main$ и следующую обобщённую кучу для $inc$:

$$Body(inc) = merge(\langle LI(node) = 0, \epsilon\rangle, \langle LI(node) \neq 0, h_1 \circ Rec(inc)\rangle),$$

где $h_1$ и $h_2$ — это определённые кучи, полученные исполнением строки 11 и строк 17-25 соответственно.

Каждая определённая куча транслируется в две функции, по одной на соответствующий тип локации. В результате программа транслируется композиционально: $inc$ соответствует той же функции на Haskell, как и $main$; если появится ещё одна функция, вызывающая $inc$, готовая Haskell версия функции $inc$ может быть переиспользована. В программах, полученных в результате такой трансляции, всегда будет только хвостовая рекурсия.

[5] Эта трансляция проведена человеком, однако менее читаемая версия может быть получена полностью автоматически

## B. Доказательства теорем

**Теорема 1.** Если σ и σ′ — символьные кучи, то σ ∘ σ′ также символьная куча.

*Доказательство.* Необходимо показать, что $\sigma \circ \sigma'$ удовлетворяет инварианту кучи (1). Рассмотрим $x, y \in dom(\sigma \circ \sigma')$.

$$union\langle x = y, (\sigma \circ \sigma')(x)\rangle \overset{\text{Опр.5 и Св.1(c)}}{=}$$
$$= union(\{\langle x = y \land x = \sigma \bullet l, \sigma \bullet (\sigma'(l))\rangle | l \in dom(\sigma')\}$$
$$\cup \langle x = y \land \bigwedge_{l \in dom(\sigma')} x \neq \sigma \bullet l, \sigma(x)\rangle) =$$
$$= union(\{\langle x = y \land x = \sigma \bullet l, \sigma \bullet (\sigma'(l))\rangle | l \in dom(\sigma')\}$$
$$\cup \langle x = y \land \bigwedge_{l \in dom(\sigma')} x \neq \sigma \bullet l, union(\langle x = y, \sigma(x)\rangle)\rangle) =$$
$$= union(\{\langle x = y \land y = \sigma \bullet l, \sigma \bullet (\sigma'(l))\rangle | l \in dom(\sigma')\}$$
$$\cup \langle x = y \land \bigwedge_{l \in dom(\sigma')} y \neq \sigma \bullet l, union(\langle x = y, \sigma(y)\rangle)\rangle) =$$
$$= union\langle x = y, (\sigma \circ \sigma')(y)\rangle$$

**Теорема 2.** Для произвольной символьной кучи $\sigma$, локации $x$ и выражения $e$ справедливо следующее:

(e) $read(\epsilon, x) = LI(x)$

(f) $\epsilon \bullet e = e$

(g) $\epsilon \circ \sigma = \sigma$

(h) $\sigma \circ \epsilon = \sigma$

*Доказательство.* Напомним, что конъюнкция пустого множества формул есть тождественная истина.

(i) $read(\epsilon, x) \overset{\text{Опр.2,3}}{=} union(\varnothing \cup \langle \top, LI(x)\rangle) \overset{\text{Св.1(a)}}{=} LI(x)$

(j) Это свойство может быть доказано структурной индукцией по выражению $e$ по опр. 4 и (a).

(k) *Во-первых,* $dom(\epsilon \circ \sigma) \overset{[0pt]\,(3)}{=} \varnothing \cup \epsilon \bullet dom(\sigma) \overset{(3)2}{=} dom(\sigma)$. *Рассмотрим* $x \in dom(\sigma)$.

$$(\epsilon \circ \sigma)(x) \overset{\text{Опр.5}}{=} union(\{\langle x = \epsilon \bullet l, \epsilon \bullet (\sigma(l))\rangle | l \in dom(\sigma)\}$$
$$\cup \langle \bigwedge_{l \in dom(\sigma)} x \neq \epsilon \bullet l, \epsilon(x)\rangle) \overset{(b)}{=}$$
$$= union(\{\langle x = l, \sigma(l)\rangle | l \in dom(\sigma)\} \cup \langle \bot, \epsilon(x)\rangle)$$
$$\overset{\text{Св.1(b)}}{=} \sigma(x)$$

(l) $dom(\sigma \circ \epsilon) \overset{(3)}{=} dom(\sigma) \cup \sigma \bullet \varnothing = dom(\sigma)$. *Пусть* $x \in dom(\sigma)$.

$$(\sigma \circ \epsilon)(x) \overset{\text{Опр.5}}{=} union(\varnothing \cup \langle \bigwedge_{l \in \varnothing} x \neq \sigma \bullet l, \sigma(x)\rangle) = \sigma(x)$$

Для доказательства следующих теорем воспользуемся схожестью структуры чтения и композиции, отражённых в определении $find$.

$$read(\sigma, x) = find(\sigma, x, \epsilon, LI(x))$$

$$(\sigma \circ \sigma')(x) = find(\sigma', x, \sigma, \sigma(x))$$

Третий аргумент $find$, $\tau$, мы будем называть *контекстной кучей*. Докажем фундаментальное свойство $find$, которое будет полезно для доказательства важных свойств композиции и уточнения:

**Лемма 1.** *Для всех таких символьных куч* $\sigma$, $\sigma'$, $\tau$, *что для каждого символьного выражения* $e$, $(\tau \circ \sigma) \bullet e = \tau \bullet (\sigma \bullet e)$, (11) и всех символьных выражений $x \in loc$, $d \in term$,

$$find(\sigma \circ \sigma', x, \tau, d) = find(\sigma', x, \tau \circ \sigma, find(\sigma, x, \tau, d))$$

*Доказательство.*

$find(\sigma \circ \sigma', x, \tau, d) =$

$= union(\{\langle x = \tau \bullet l, \tau \bullet (\sigma \circ \sigma')(l)\rangle | l \in dom(\sigma \circ \sigma')\} \cup$

$\cup \{\langle \bigwedge_{l \in dom(\sigma \circ \sigma')} x \neq \tau \bullet l, d\rangle\}) =$

$= union(\{\langle x = \tau \bullet l, \tau \bullet union(\{\langle l = \sigma \bullet l', \sigma \bullet (\sigma'(l'))\rangle | l' \in dom(\sigma')\} \cup$

$\cup \{\langle \bigwedge_{l' \in dom(\sigma')} l \neq \sigma \bullet l', \sigma(l)\rangle\})\rangle | l \in dom(\sigma \circ \sigma')\} \cup \{\langle \bigwedge_{l \in dom(\sigma \circ \sigma')} x$

$\neq \tau \bullet l, d\rangle\}) \overset{\text{Св.1.}(c)}{=}$

$= union(\{\langle x = \tau \bullet l \wedge \tau \bullet l = \tau \bullet (\sigma \bullet l'), \tau \bullet (\sigma \bullet (\sigma'(l')))\rangle$

$| l' \in dom(\sigma'), l \in dom(\sigma \circ \sigma')\} \cup$

$\cup \{\langle x = \tau \bullet l \wedge \bigwedge_{l' \in dom(\sigma')} \tau \bullet l \neq \tau \bullet (\sigma \bullet l'), \tau \bullet (\sigma(l))\rangle | l \in dom(\sigma \circ \sigma')\} \cup$

$\cup \{\langle \bigwedge_{l \in dom(\sigma \circ \sigma')} x \neq \tau \bullet l, d\rangle\}) \overset{(3)}{=}$

$= union(\{\langle x = \tau \bullet l \wedge \tau \bullet l = \tau \bullet (\sigma \bullet l'), \tau \bullet (\sigma \bullet (\sigma'(l')))\rangle$

$| l' \in dom(\sigma'), l \in dom(\sigma) \cup \sigma \bullet dom(\sigma')\} \cup$

$\cup \{\langle x = \tau \bullet l \wedge \bigwedge_{l' \in dom(\sigma')} \tau \bullet l \neq \tau \bullet (\sigma \bullet l'), \tau \bullet (\sigma(l))\rangle$

$| l \in dom(\sigma) \cup \sigma \bullet dom(\sigma')\} \cup \{\langle \bigwedge_{l \in dom(\sigma) \cup \sigma \bullet dom(\sigma')} x \neq \tau \bullet l, d\rangle\}) =$

$= union(\{\langle x = \tau \bullet l \wedge \tau \bullet l = \tau \bullet (\sigma \bullet l'), \tau \bullet (\sigma \bullet (\sigma'(l')))\rangle$

$| l' \in dom(\sigma'), l \in \sigma \bullet dom(\sigma')\} \cup$

$\cup \{\langle x = \tau \bullet l \wedge \bigwedge_{l' \in dom(\sigma')} \tau \bullet l \neq \tau \bullet (\sigma \bullet l'), \tau \bullet (\sigma(l))\rangle$

$| l \in dom(\sigma)\} \cup \{\langle \bigwedge_{l \in dom(\sigma) \cup \sigma \bullet dom(\sigma')} x \neq \tau \bullet l, d\rangle\}) =$

$= union(\{\langle x = \tau \bullet (\sigma \bullet l'), \tau \bullet (\sigma \bullet (\sigma'(l')))\rangle | l' \in dom(\sigma')\} \cup$

$\cup \{\langle x = \tau \bullet l \wedge \bigwedge_{l' \in dom(\sigma')} x \neq \tau \bullet (\sigma \bullet l'), \tau \bullet (\sigma(l))\rangle | l \in dom(\sigma)\}$

$\cup \langle \bigwedge_{l' \in dom(\sigma')} x \neq \tau \bullet (\sigma \bullet l') \wedge \bigwedge_{l \in dom(\sigma)} x \neq \tau \bullet l, d\rangle) \overset{(11)}{=}$

$= union(\{\langle x = (\tau \circ \sigma) \bullet l', (\tau \circ \sigma) \bullet (\sigma'(l'))\rangle | l' \in dom(\sigma')\} \cup$

$\cup \{\langle x = \tau \bullet l \wedge \bigwedge_{l' \in dom(\sigma')} x \neq (\tau \circ \sigma) \bullet l', \tau \bullet (\sigma(l))\rangle | l \in dom(\sigma)\}$

$$\cup \langle \bigwedge_{l' \in dom(\sigma')} x \neq (\tau \circ \sigma) \bullet l' \wedge \bigwedge_{l \in dom(\sigma)} x \neq \tau \bullet l, d \rangle) \overset{\text{Св.1(c)}}{=}$$

$$= union(\{\langle x = (\tau \circ \sigma) \bullet l', (\tau \circ \sigma) \bullet (\sigma'(l')) \rangle | l' \in dom(\sigma')\}$$

$$\cup \langle \bigwedge_{l' \in dom(\sigma')} x \neq (\tau \circ \sigma) \bullet l', find(\sigma, x, \tau, d) \rangle) =$$

$$= find(\sigma', x, \tau \circ \sigma, find(\sigma, x, \tau, d))$$

**Утверждение 3.** Для всех определённых символьных куч $\sigma$, $\sigma'$, $\tau$, таких что для каждого символьного выражения $e$, выполняется $(\tau \circ \sigma) \bullet e = \tau \bullet (\sigma \bullet e)$, и всех символьных выражений $x \in loc$, $d \in term$,

$$\tau \bullet find(\sigma', x, \sigma, d) = find(\sigma', \tau \bullet x, \tau \circ \sigma, \tau \bullet d).$$

*Доказательство.*

$$\tau \bullet find(\sigma', x, \sigma, d) =$$

$$\tau \bullet union(\{\langle x = \sigma \bullet l, \sigma \bullet (\sigma'(l)) \rangle | l \in dom(\sigma')\} \cup \langle \bigwedge_{l \in dom(\sigma')} x \neq \sigma \bullet l, d \rangle) \overset{\text{Опр.4}}{=}$$

$$= union(\{\langle \tau \bullet x = \tau \bullet (\sigma \bullet l), \tau \bullet (\sigma \bullet (\sigma'(l))) \rangle | l \in dom(\sigma')\}$$

$$\cup \{\langle \bigwedge_{l \in dom(\sigma')} \tau \bullet x \neq \tau \bullet (\sigma \bullet l), \tau \bullet d \rangle\}) \overset{(11)}{=}$$

$$= union(\{\langle \tau \bullet x = (\tau \circ \sigma) \bullet l, (\tau \circ \sigma) \bullet (\sigma'(l)) \rangle | l \in dom(\sigma')\}$$

$$\cup \{\langle \bigwedge_{l \in dom(\sigma')} \tau \bullet x \neq (\tau \circ \sigma) \bullet l, \tau \bullet d \rangle\}) =$$

$$= find(\sigma', \tau \bullet x, \tau \circ \sigma, \tau \bullet d)$$

**Теорема 3.** Для всех символьных куч $\sigma$, $\sigma'$ и символьных локаций $x$ справедливо следующее:

$$\sigma \bullet read(\sigma', x) = read(\sigma \circ \sigma', \sigma \bullet x).$$

*Доказательство.* Заметим, что по свойствам теор. 2, (11), выполняющимся для $\sigma = \epsilon$ и $\tau = \epsilon$:

$$\sigma \bullet read(\sigma', x) \overset{(4)}{=} \sigma \bullet find\big(\sigma', x, \epsilon, LI(x)\big)$$

$$\overset{\text{утв.3}}{=} find(\sigma', \sigma \bullet x, \sigma \circ \epsilon, \sigma \bullet LI(x)) \overset{\text{Теор.2,Опр.4}}{=}$$

$$= find\big(\sigma', \sigma \bullet x, \sigma, read(\sigma, \sigma \bullet x)\big)$$

$$\overset{(4)}{=} find(\sigma', \sigma \bullet x, \sigma, find(\sigma, \sigma \bullet x, \epsilon, LI(\sigma \bullet x))) \overset{\text{Лем.1}}{=}$$

$$= find(\sigma \circ \sigma', \sigma \bullet x, \epsilon, LI(\sigma \bullet x)) \overset{(4)}{=} read(\sigma \circ \sigma', \sigma \bullet x)$$

**Теорема 4.** Для всех символьных куч $\sigma$, $\sigma'$ и символьных выражений $e$ справедливо следующее:

$$(\sigma \circ \sigma') \bullet e = \sigma \bullet (\sigma' \bullet e).$$

*Доказательство.* Доказательство структурной индукцией по опр. 4. Случаи 1-3 тривиальны. Единственное, что необходимо показать, это $(\sigma \circ \sigma') \bullet LI(x) = \sigma \bullet (\sigma' \bullet LI(x))$. Индукционная гипотеза выглядит так:

$$(\sigma \circ \sigma') \bullet x = \sigma \bullet (\sigma' \bullet x)$$

Теперь

$$\sigma \bullet (\sigma' \bullet LI(x)) \overset{\text{Опр.4}}{=} \sigma \bullet read(\sigma', \sigma' \bullet x) \overset{\text{Теор.3}}{=} read(\sigma \circ \sigma', \sigma \bullet (\sigma' \bullet x)) \overset{I.H.}{=}$$
$$= read(\sigma \circ \sigma', (\sigma \circ \sigma') \bullet x) \overset{\text{Опр.4}}{=} (\sigma \circ \sigma') \bullet LI(x)$$

**Лемма 2.** *Для всех символьных куч $\sigma$, $\sigma'$ и локаций $x$,*

$$read(\sigma \circ \sigma', x) = find(\sigma', x, \sigma, read(\sigma, x))$$

*Доказательство.*

$$read(\sigma \circ \sigma', x) = find\big(\sigma \circ \sigma', x, \epsilon, LI(x)\big) =$$
$$\overset{\text{Лем.1}}{=} find(\sigma', x, \epsilon \circ \sigma, find(\sigma, x, \epsilon, LI(x))) = find(\sigma', x, \sigma, read(\sigma, x))$$

**Теорема 5.** Для всех символьных куч $\sigma_1$, $\sigma_2$ и $\sigma_3$ справедливо следующее:

$$(\sigma_1 \circ \sigma_2) \circ \sigma_3 = \sigma_1 \circ (\sigma_2 \circ \sigma_3).$$

*Доказательство.* Для начала покажем равенство их областей действия.

$$dom((\sigma_1 \circ \sigma_2) \circ \sigma_3) \overset{(3)}{=} dom(\sigma_1 \circ \sigma_2) \cup \{(\sigma_1 \circ \sigma_2) \bullet l_3 | l_3 \in dom(\sigma_3)\} \overset{(3)}{=}$$
$$= dom(\sigma_1) \cup \{\sigma_1 \bullet l_2 | l_2 \in dom(\sigma_2)\} \cup \{(\sigma_1 \circ \sigma_2) \bullet l_3 | l_3 \in dom(\sigma_3)\} \overset{\text{теор.4}}{=}$$
$$= dom(\sigma_1) \cup \{\sigma_1 \bullet l_2 | l_2 \in dom(\sigma_2)\} \cup \{\sigma_1 \bullet (\sigma_2 \bullet l_3) | l_3 \in dom(\sigma_3)\} =$$
$$= dom(\sigma_1) \cup \{\sigma_1 \bullet l_2 | l_2 \in dom(\sigma_2)\} \cup \{\sigma_1 \bullet l_{3'} | l_{3'} \in \sigma_2 \bullet dom(\sigma_3)\} =$$
$$= dom(\sigma_1) \cup \{\sigma_1 \bullet l_{23} | l_{23} \in dom(\sigma_2) \cup \sigma_2 \bullet dom(\sigma_3)\} =$$
$$= dom(\sigma_1) \cup \{\sigma_1 \bullet l_{23} | l_{23} \in dom(\sigma_2 \circ \sigma_3)\} \overset{(3)}{=} dom(\sigma_1 \circ (\sigma_2 \circ \sigma_3))$$

Рассмотрим произвольное $x \in dom((\sigma_1 \circ \sigma_2) \circ \sigma_3)$,

$$((\sigma_1 \circ \sigma_2) \circ \sigma_3)(x) = find(\sigma_3, x, \sigma_1 \circ \sigma_2, find(\sigma_2, x, \sigma_1, \sigma_1(x))) \overset{\text{лем.1}}{=}$$
$$= find(\sigma_2 \circ \sigma_3, x, \sigma_1, \sigma_1(x)) = (\sigma_1 \circ (\sigma_2 \circ \sigma_3))(x)$$

Для упрощения внешнего вида доказательств, мы сокращаем нотацию $merge(\langle g_1, \sigma_1 \rangle, \dots, \langle g_n, \sigma_n \rangle)$ и пишем просто $merge\langle g_i, \sigma_i \rangle$; аналогично вместо $union(\langle g_1, v_1 \rangle, \dots, \langle g_n, v_n \rangle)$ далее будем писать $union\langle g_i, v_i \rangle$.

**Теорема 7.** Для любой символьной кучи $\sigma$ и символьных локаций $x$, $y$ справедливо следующее:

$$union(\langle x = y, read(\sigma, x) \rangle) = union(\langle x = y, read(\sigma, y) \rangle).$$

*Доказательство.*

$$union\langle x = y, read(\sigma, x) \rangle =$$
$$= union(\langle x = y, union(\{\langle x = l, \sigma(l) \rangle | l \in dom(\sigma)\} \cup \langle \bigwedge_{l \in dom(\sigma)} x \neq l, LI(x) \rangle) \rangle) =$$
$$= union(\{\langle x = y \wedge x = l, \sigma(l) \rangle | l \in dom(\sigma)\} \cup \langle x = y \wedge \bigwedge_{l \in dom(\sigma)} x \neq l, LI(x) \rangle) =$$
$$= union(\{\langle x = y \wedge y = l, \sigma(l) \rangle | l \in dom(\sigma)\} \cup \langle x = y \wedge \bigwedge_{l \in dom(\sigma)} y \neq l, LI(y) \rangle)$$
$$= union\langle x = y, read(\sigma, y) \rangle$$

**Теорема 8.** Для всех символьных куч $\sigma_1, \dots, \sigma_n$ и непересекающихся ограничений $g_1, \dots, g_n$, $merge\langle g_i, \sigma_i \rangle$ — символьная куча.

*Доказательство.* Возьмём $x, y \in dom(merge\langle g_i, \sigma_i \rangle) = \bigcup_{i=1}^{n} dom(\sigma_i)$.

$$union\langle x = y, merge\langle g_i, \sigma_i \rangle(x) \rangle = union\langle x = y, union\langle g_i, read(\sigma_i, x) \rangle \rangle =$$

$$= union\langle x = y \wedge g_i, read(\sigma_i, x)\rangle \overset{\text{теор.7}}{=} union\langle x = y \wedge g_i, read(\sigma_i, y)\rangle =$$
$$= union\langle x = y, merge\langle g_i, \sigma_i\rangle(y)\rangle$$

**Лемма 3.** *Для всех символьных куч* $\sigma_1, \dots, \sigma_n, \tau$, *непересекающихся ограничений* $g_1, \dots, g_n$, символьных локаций $x$ и символьных выражений $d$,

$$find(merge\langle g_i, \sigma_i\rangle, x, \tau, read(\tau, x)) = union\langle \tau \bullet g_i, find(\sigma_i, x, \tau, read(\tau, x))\rangle$$

*Доказательство.* Сначала заметим, что для любого предиката $p$ и непересекающихся множеств $A, B$ верно

$$\bigvee_{u \in A \coprod B} (p(u) \wedge \bigwedge_{v \in B} \neg p(v)) \vee \bigwedge_{u \in A \coprod B} \neg p(u) \Leftarrow \bigwedge_{u \in B} \neg p(v) \qquad (12)$$

Рассмотрим произвольную символьную кучу $\sigma$ и произвольное множество символьных локаций $A$. Можно допустить, что $A$ и $dom(\sigma)$ не пересекаются, иначе возьмём $A := A \backslash dom(\sigma)$. Теперь

$$union(\{\langle x = \tau \bullet l, \tau \bullet read(\sigma, l)\rangle | l \in dom(\sigma) \cup A\}$$
$$\cup \langle \bigwedge_{l \in dom(\sigma) \cup A} x \neq \tau \bullet l, read(\tau, x)\rangle) =$$
$$= union(\{\langle x = \tau \bullet l, union(\{\langle \tau \bullet l = \tau \bullet l', \tau \bullet (\sigma(l'))\rangle | l' \in dom(\sigma)\} \cup$$
$$\cup \langle \bigwedge_{l' \in dom(\sigma)} \tau \bullet l \neq \tau \bullet l', \tau \bullet LI(l)\rangle)\rangle | l \in dom(\sigma) \cup A\} \cup$$
$$\cup \langle \bigwedge_{l \in dom(\sigma) \cup A} x \neq \tau \bullet l, read(\tau, x)\rangle) =$$
$$= union(\{\langle x = \tau \bullet l \wedge \tau \bullet l = \tau \bullet l', \tau \bullet (\sigma(l'))\rangle | l \in dom(\sigma) \cup A, l' \in dom(\sigma)\} \cup$$
$$\cup \{\langle x = \tau \bullet l \wedge \bigwedge_{l' \in dom(\sigma)} \tau \bullet l \neq \tau \bullet l', read(\tau, \tau \bullet l)\rangle$$
$$| l \in dom(\sigma) \cup A\} \cup \langle \bigwedge_{l \in dom(\sigma) \cup A} x \neq \tau \bullet l, read(\tau, x)\rangle) \overset{\text{Теор.7}}{=}$$
$$= union(\{\langle x = \tau \bullet l \wedge x = \tau \bullet l', \tau \bullet (\sigma(l'))\rangle | l \in dom(\sigma) \cup A, l' \in dom(\sigma)\} \cup$$
$$\cup \{\langle x = \tau \bullet l \wedge \bigwedge_{l' \in dom(\sigma)} x \neq \tau \bullet l', read(\tau, x)\rangle | l \in dom(\sigma) \cup A\} \cup$$
$$\cup \langle \bigwedge_{l \in dom(\sigma) \cup A} x \neq \tau \bullet l, read(\tau, x)\rangle) \overset{\text{Св.1}(d)}{=}$$
$$= union(\{\langle x = \tau \bullet l', \tau \bullet (\sigma(l'))\rangle | l' \in dom(\sigma)\} \cup$$
$$\cup \langle \bigvee_{l \in dom(\sigma) \cup A} (x = \tau \bullet l \wedge \bigwedge_{l' \in dom(\sigma)} x \neq \tau \bullet l') \vee \bigwedge_{l \in dom(\sigma) \cup A} x$$
$$\neq \tau \bullet l, read(\tau, x)\rangle) \overset{(12)}{=}$$
$$= union(\{\langle x = \tau \bullet l, \tau \bullet (\sigma(l))\rangle | l \in dom(\sigma)\} \cup \langle \bigwedge_{l \in dom(\sigma)} x \neq \tau \bullet l, read(\tau, x)\rangle) =$$
$$= find(\sigma, x, \tau, read(\tau, x)) \qquad (13)$$

Наконец,

$$find(merge\langle g_i, \sigma_i\rangle, x, \tau, read(\tau, x)) =$$
$$= union(\{\langle x = \tau \bullet l, \tau \bullet union\langle g_i, read(\sigma_i, l)\rangle\rangle | l \in \bigcup_{j=1}^{n} dom(\sigma_j)\} \cup$$
$$\cup \langle \bigwedge_{l \in \bigcup_{j=1}^{n} dom(\sigma_j)} x \neq \tau \bullet l, read(\tau, x)\rangle) =$$

$$= union\langle x = \tau \bullet g_i, union(\{\langle x = \tau \bullet l, \tau \bullet read(\sigma_i, l)\rangle \mid l \in \bigcup_{j=1}^{n} dom(\sigma_j)\} \cup$$

$$\cup \langle \bigwedge_{l \in \bigcup_{j=1}^{n} dom(\sigma_j)} x \neq \tau \bullet l, read(\tau, x)\rangle)\rangle \overset{(13)}{=}$$

$$= union\langle \tau \bullet g_i, find(\sigma_i, x, \tau, read(\tau, x))\rangle$$

***Теорема 9.*** *Для всех символьных куч $\sigma_1, \dots, \sigma_n$, непересекающихся ограничений $g_1, \dots, g_n$ и локаций $x$ справедливо следующее:*

$$read(merge\langle g_i, \sigma_i\rangle, x) = union\langle g_i, read(\sigma_i, x)\rangle.$$

*Доказательство.*

$$read(merge\langle g_i, \sigma_i\rangle, x) = find(merge\langle g_i, \sigma_i\rangle, x, \epsilon, read(\epsilon, x)) =$$

$$\overset{\text{лем.3}}{=} union\langle \epsilon \bullet g_i, find(\sigma_i, x, \epsilon, read(\epsilon, x))\rangle = union\langle g_i, read(\sigma_i, x)\rangle$$

***Теорема 10.*** *Для любых символьных куч $\sigma, \sigma_1, \dots, \sigma_n$ и непересекающихся ограничений $g_1, \dots, g_n$ выполняется следующее утверждение:*

$$\sigma \circ merge(\langle g_1, \sigma_1\rangle, \dots, \langle g_n, \sigma_n\rangle) = merge(\langle \sigma \bullet g_1, \sigma \circ \sigma_1\rangle, \dots, \langle \sigma \bullet g_n, \sigma \circ \sigma_n\rangle).$$

*Доказательство.*

$$dom(\sigma \circ merge(\langle g_1, \sigma_1\rangle, \dots, \langle g_n, \sigma_n\rangle)) =$$

$$= dom(\sigma) \cup \sigma \bullet dom(merge(\langle g_1, \sigma_1\rangle, \dots, \langle g_n, \sigma_n\rangle)) =$$

$$= dom(\sigma) \cup \sigma \bullet \bigcup_{i=0}^{n} dom(\sigma_i) = \bigcup_{i=0}^{n} (dom(\sigma) \cup \sigma \bullet dom(\sigma_i)) =$$

$$= \bigcup_{i=0}^{n} dom(\sigma \circ \sigma_i) = dom(merge(\langle \sigma \bullet g_1, \sigma \circ \sigma_1\rangle, \dots, \langle \sigma \bullet g_n, \sigma \circ \sigma_n\rangle))$$

Возьмём $x \in dom(\sigma \circ merge(\langle g_1, \sigma_1\rangle, \dots, \langle g_n, \sigma_n\rangle))$. Напомним, что для $x \in dom(\sigma), \sigma(x) = read(\sigma, x)$.

$$(\sigma \circ merge(\langle g_1, \sigma_1\rangle, \dots, \langle g_n, \sigma_n\rangle))(x) =$$

$$= read(\sigma \circ merge(\langle g_1, \sigma_1\rangle, \dots, \langle g_n, \sigma_n\rangle), x) \overset{\text{лем.2}}{=}$$

$$= find(merge(\langle g_1, \sigma_1\rangle, \dots, \langle g_n, \sigma_n\rangle), x, \sigma, read(\sigma, x)) \overset{\text{лем.3}}{=}$$

$$= union(\langle \sigma \bullet g_1, find(\sigma_1, x, \sigma, read(\sigma, x))\rangle, \dots,$$

$$\langle \sigma \bullet g_n, find(\sigma_n, x, \sigma, read(\sigma, x))\rangle) \overset{\text{лем.2}}{=}$$

$$= union(\langle \sigma \bullet g_1, read(\sigma \circ \sigma_1, x)\rangle, \dots, \langle \sigma \bullet g_n, read(\sigma \circ \sigma_n, x)\rangle) \overset{\text{лем.2}}{=}$$

$$= (merge(\langle \sigma \bullet g_1, \sigma \circ \sigma_1\rangle, \dots, \langle \sigma \bullet g_n, \sigma \circ \sigma_n\rangle))(x)$$

***Теорема 11.*** *Для всех символьных куч $\sigma$, непересекающихся ограничений $g_1, \dots, g_n$ и локаций $x_1, \dots, x_n$ справедливо следующее:*

$$read(\sigma, union\langle g_i, x_i\rangle) = union\langle g_i, read(\sigma, x_i)\rangle.$$

*Доказательство.*

$$read(\sigma, union\langle g_i, x_i\rangle) =$$

$$= union(\{\langle union\langle g_i, x_i\rangle = l, \sigma(l)\rangle | l \in dom(\sigma)\} \cup \langle \bigwedge_{l\in dom(\sigma)} union\langle g_i, x_i\rangle \neq l, LI(union\langle g_i, x_i\rangle)\rangle) \overset{\text{Св.1(6)}}{=}$$

$$= union\langle g_i, union(\{\langle x_i = l, \sigma(l)\rangle | l \in dom(\sigma)\} \cup \langle \bigwedge_{l\in dom(\sigma)} x_i \neq l, LI(x_i)\rangle)\rangle = union\langle g_i, read(\sigma, x_i)\rangle$$

***Теорема 12.*** *Для всех символьных куч* $\sigma_1, \dots, \sigma_n$, *непересекающихся ограничений* $g_1, \dots, g_n$ *и выражений е справедливо следующее:*

$$union\langle g_1 \vee \dots \vee g_n, merge\langle g_i, \sigma_i\rangle \bullet e\rangle = union\langle g_i, \sigma_i \bullet e\rangle.$$

*Доказательство.* Доказательство структурной индукцией по опр. 4. Возьмём $G \overset{\text{def}}{=} g_1 \vee \dots \vee g_n$ и $M \overset{\text{def}}{=} merge\langle g_i, \sigma_i\rangle$.

(m) $union\langle G, M \bullet leaf\rangle = union\langle G, leaf\rangle \overset{\text{Св.1.(c)}}{=} union\langle g_i, leaf\rangle = union\langle g_i, \sigma_i \bullet leaf\rangle$

(n) $union\langle G, M \bullet op(e_1, \dots, e_m)\rangle = union\langle G, op(M \bullet e_1, \dots, M \bullet e_m)\rangle \overset{I.H.}{=}$

$$= union\langle G, op(union\langle g_i, \sigma_i \bullet e_1\rangle, \dots, union\langle g_i, \sigma_i \bullet e_m\rangle)\rangle \overset{\text{Св.1(c,e)}}{=}$$

$$= op(union\langle g_i, \sigma_i \bullet e_1\rangle, \dots, union\langle g_i, \sigma_i \bullet e_m\rangle) \overset{\text{Св.1(e)}}{=} union\langle g_i, \sigma_i \bullet op(e_1, \dots, e_m)\rangle$$

(o) *То же, что в (b)*

(p) $union\langle G, M \bullet LI(x)\rangle = union\langle G, read(M, M \bullet x)\rangle \overset{\text{Теор.9}}{=}$

$$= union\langle G, union\langle g_i, read(\sigma_i, M \bullet x)\rangle\rangle \overset{\text{Св.1(c)}}{=} union\langle g_i, read(\sigma_i, M \bullet x)\rangle \overset{I.H.}{=}$$

$$= union\langle g_i, read(\sigma_i, union\langle g_j, \sigma_j \bullet x\rangle)\rangle \overset{\text{Теор.11}}{=}$$

$$union\langle g_i, union\langle g_j, read(\sigma_i, \sigma_j \bullet x)\rangle\rangle \overset{g_i\ \text{непересек.}}{=}$$

$$= union\langle g_i, read(\sigma_i, \sigma_i \bullet x)\rangle = union\langle g_i, \sigma_i \bullet LI(x)\rangle$$

***Теорема 13.*** *Для всех символьных куч* $\sigma, \sigma_1, \dots, \sigma_n$, *непересекающихся ограничений* $g_1, \dots, g_n$ *и локаций x справедливо следующее:*

$$union\langle g_1 \vee \dots \vee g_n, read(merge\langle g_i, \sigma_i\rangle \circ \sigma, x)\rangle = read(merge\langle g_i, \sigma_i \circ \sigma\rangle, x).$$

*Доказательство.* Пусть $G \overset{\text{def}}{=} g_1 \vee \dots \vee g_n$ и $M \overset{\text{def}}{=} merge\langle g_i, \sigma_i\rangle$.

$$union\langle G, read(M \circ \sigma, x)\rangle \overset{\text{лем.2}}{=} union\langle G, find(\sigma, x, M, read(M, x))\rangle =$$

$$= union\langle G union(\langle x = M \bullet l, M \bullet (\sigma(l))\rangle | l \in dom(\sigma)$$

$$\cup \langle \bigwedge_{l\in dom(\sigma)} x \neq M \bullet l, read(M, x)\rangle)\rangle \overset{\text{Теор 12}}{=}$$

$$= union\langle g_i union(\langle x = \sigma_i \bullet l, \sigma_i \bullet (\sigma(l))\rangle | l \in dom(\sigma)$$

$$\cup \langle \bigwedge_{l\in dom(\sigma)} x \neq \sigma_i \bullet l, read(\sigma_i, x)\rangle)\rangle =$$

$$= union\langle g_i, find(\sigma, x, \sigma_i, read(\sigma_i, x))\rangle \overset{\text{Лем.2}}{=}$$

$$= union\langle g_i, read(\sigma_i \circ \sigma, x)\rangle = read(merge\langle g_i, \sigma_i \circ \sigma\rangle, x)$$

***Лемма 4*** *Для всех символьных куч* $\sigma$, $\tau$, *символьных локаций* $x$, $y$ *и символьных выражений* $v$, $d$,

$$find(write(\sigma, y, v), x, \tau, d) = ite(x = \tau \bullet y, \tau \bullet v, find(\sigma, x, \tau, d))$$

*Доказательство.*

$$find(write(\sigma, y, v), x, \tau, d) =$$
$$= union(\{\langle x = \tau \bullet l, \tau \bullet (write(\sigma, y, v)(l))\rangle | l \in dom(write(\sigma, y, v))\}$$
$$\cup \langle \bigwedge_{l \in dom(write(\sigma,y,v))} x \neq \tau \bullet l, d\rangle) =$$
$$= union(\{\langle x = \tau \bullet l, \tau \bullet ite(y = l, v, \sigma(l))\rangle | l \in dom(\sigma) \cup \{y\}\} \cup$$
$$\cup \langle \bigwedge_{l \in dom(\sigma) \cup \{y\}} x \neq \tau \bullet l, d\rangle) =$$
$$= union(\{\langle x = \tau \bullet l, ite(\tau \bullet y = \tau \bullet l, \tau \bullet v, \tau \bullet (\sigma(l)))\rangle | l \in dom(\sigma)\} \cup$$
$$\cup \langle x = \tau \bullet y, \tau \bullet v\rangle \cup \langle \bigwedge_{l \in dom(\sigma)} x \neq \tau \bullet l \wedge x \neq \tau \bullet y, d\rangle) =$$
$$= union(\{\langle x = \tau \bullet l \wedge \tau \bullet y = \tau \bullet l, \tau \bullet v\rangle | l \in dom(\sigma)\} \cup$$
$$\cup \{\langle x = \tau \bullet l \wedge \tau \bullet y \neq \tau \bullet l, \tau \bullet (\sigma(l))\rangle | l \in dom(\sigma)\} \cup$$
$$\cup \langle x = \tau \bullet y, \tau \bullet v\rangle \cup \langle \bigwedge_{l \in dom(\sigma)} x \neq \tau \bullet l \wedge x \neq \tau \bullet y, d\rangle) =$$
$$= ite(x = \tau \bullet y, \tau \bullet v union(\langle x = \tau \bullet l, \tau \bullet (\sigma(l))\rangle | l \in dom(\sigma)$$
$$\cup \langle \bigwedge_{l \in dom(\sigma)} x \neq \tau \bullet l, d\rangle)) =$$
$$= ite(x = \tau \bullet y, \tau \bullet v, find(\sigma, x, \tau, d))$$

***Теорема 14.*** *Для всех символьных куч* $\sigma$, *символьных локаций* $x$, $y$ *и символьных выражений* $v$ *справедливо следующее:*

$$read(write(\sigma, y, v), x) = ite(x = y, v, read(\sigma, x)).$$

*Доказательство.*

$$read(write(\sigma, y, v), x) = find(write(\sigma, y, v), x, \epsilon, LI(x)) \overset{\text{Лем.4}}{=}$$
$$= ite(x = \epsilon \bullet y, \epsilon \bullet v, find(\sigma, x, \epsilon, LI(x))) = ite(x = y, v, read(\sigma, x))$$

***Теорема 15.*** *Для всех символьных куч* $\sigma$, $\sigma'$, *символьных локаций* $y$ *и символьных выражений* $v$ *справедливо следующее:*

$$\sigma \circ write(\sigma', y, v) = write(\sigma \circ \sigma', \sigma \bullet y, \sigma \bullet v).$$

*Доказательство.*

$$(\sigma \circ write(\sigma', y, v))(x) = find(write(\sigma', y, v), x, \sigma, \sigma(x)) \overset{\text{Лем.4}}{=}$$
$$= ite(x = \sigma \bullet y, \sigma \bullet v, find(\sigma', x, \sigma, \sigma(x))) =$$
$$= ite(x = \sigma \bullet y, \sigma \bullet v, (\sigma \circ \sigma')(x)) = write(\sigma \circ \sigma', \sigma \bullet y, \sigma \bullet v)$$

***Теорема 16.*** *Для всех символьных куч* $\sigma_1, \dots, \sigma_n$, *непересекающихся ограничений* $g_1, \dots, g_n$, *символьных локаций* $y$ *и символьных выражений* $v$ *справедливо следующее:*

$$write(merge\langle g_i, \sigma_i\rangle, y, v) = merge\langle g_i, write(\sigma_i, y, v)\rangle.$$

*Доказательство.*

$$dom(write(merge\langle g_i, \sigma_i\rangle, y, v)) = \bigcup_{i=1}^{n} dom(\sigma_i) \cup \{y\} =$$
$$= dom(merge\langle g_i, write(\sigma_i, y, v)\rangle)$$

Пусть $x \in dom(write(merge\langle g_i, \sigma_i\rangle, y, v))$.

$$write(merge\langle g_i, \sigma_i\rangle, y, v)(x) = ite(x = y, v, merge\langle g_i, \sigma_i\rangle(x)) =$$
$$= ite(x = y, v, union\langle g_i, read(\sigma_i, x)\rangle) \overset{\text{Св.1}}{=} union\langle g_i, ite(x = y, v, read(\sigma_i, x))\rangle$$
$$= merge\langle g_i, write(\sigma_i, y, v)\rangle(x)$$

## *C. Метод описания всех путей в CFG*

**Определение 13.** *Граф потока управления* (CFG) — это четверка *($V_G$, $E_G$, $start$, $exit$),* где:

- $V_G$ — это множество вершин графа, каждая из которых соответствует инструкции программы;
- $E_G$ — это множество ориентированных ребер $e = (u, v)$ графа, каждое ребро обозначает способность передачи управления от вершины $u$ к вершине $v$;
- $start$ — это стартовая вершина (инструкция), в которую не входит ни одно ребро графа;
- $exit$ — это финальная вершина (инструкция), из которой не выходит ни одного ребра графа.

**Определение 14.** *Путём* $p$ в графе $G$ будем называть последовательность ребер $e_1 \dots e_n$ такую, что $\forall i \ 1 \le i \le n-1$, $end(e_i) = beg(e_{i+1})$. Началом пути $beg(p)$ обозначим вершину $beg(e_1)$. Концом пути $end(p)$ обозначим вершину $end(e_n)$. Иногда удобно представлять путь в графе в виде последовательности вершин: $v_1 \to v_2 \to \dots \to v_n \to v_{n+1}$ такой, что $\forall i \ 1 \le i \le n$, $v_i = beg(e_i)$ и $v_{i+1} = end(e_i)$.

**Замечание 1.** Путь, состоящий из 0 рёбер, называется *пустым* (обозначим его $\varepsilon$).

**Определение 15.** Пусть даны два пути $p_1$ и $p_2$ в графе $G$. Если $end(p_1) = beg(p_2)$, то можно определить *конкатенацию* путей $p_1$ и $p_2$, обозначаемую $p_1 \circ p_2$ и представляющую путь, содержащий все ребра пути $p_1$, за которыми следуют ребра $p_2$. Заметим, что пустой путь $\varepsilon$ является нейтральным элементов по отношению к операции конкатенации.

Аналогично определим конкатенацию двух множеств путей.

$$P_1 \circ P_2 = \{p_1 \circ p_2 \quad | \quad p_1 \in P_1, p_2 \in P_2\}.$$

И конкатенацию пути и множества путей.

$$p_1 \circ P_2 = \{p_1\} \circ P_2.$$

**Определение 16.** Если длина пути $p = v_0 \to v_1 \to \ldots \to v_n \to v_{n+1}$ больше, либо равна 2, то будем называть серединой пути множество вершин $\{v_i | 1 \leq i \leq n\}$. Если длина пути меньше 2, то середина пути — ∅.

**Определение 17.** $\Pi(u, v, D)$ — символ для множества путей, начинающихся в вершине $u$, заканчивающихся в вершине $v$ и не проходящих через *рекурсивные* (см. опр. 10) вершины множества $D$ в середине пути. Это множество путей строится с помощью следующих правил.

- Если в графе есть ребро $(u, v)$, то путь, состоящий из одного ребра, добавляется во множество путей $\Pi(u, v, D)$ (см. правило I).
- Если есть ребро $(u, t)$, где $t \neq v$ и $t \notin RV$, то конкатенация ребра $(u, t)$ и путей $\Pi(t, v, D)$ добавляется к множеству путей $\Pi(u, v, D)$ (см. правило II).
- Если есть ребро $(u, t)$, где $t \neq v$, $t \in RV$ и $t \notin D$, то конкатенация ребра $(u, t)$ и $Rec(t, D \cup \{t\}) \circ \Pi(t, v, D \cup \{t\})$ добавляется к множеству путей $\Pi(u, v, D)$ (см. правило III), где $Rec(t, D \cup \{t\})$ — множество циклов из $t$ в $t$, не проходящих через *рекурсивные* вершины из множества $D \cup \{t\}$ в середине пути, а $\Pi(t, v, D \cup \{t\})$ — множество путей из $t$ в $v$, не проходящих через *рекурсивные* вершины из множества $D \cup \{t\}$ в середине пути.

$Rec(u, D)$ — рекурсивный символ для множества циклов из вершины $u$ в вершину $u$, не проходящих через вершины из множества $D$ в середине пути, которые построены по правилу IV.

- Пустой путь $\varepsilon$ принадлежит множеству $Rec(u, D)$.
- Конкатенация путей $\Pi(u, u, D)$ и циклов $Rec(u, D)$ добавляется к множеству циклов $Rec(u, D)$.

$$\Pi(u, v, D) = \bigcup_{(u,v) \in E_G} \{(u, v)\} \cup \quad \text{(I)}$$

$$\cup \bigcup_{\substack{t \notin RV \\ t \neq v \\ (u,t) \in E_G}} (u, t) \circ \Pi(t, v, D) \cup \quad \text{(II)}$$

$$\cup \bigcup_{\substack{t \in RV \\ t \neq v \\ (u,t) \in E_G \\ t \notin D}} (u, t) \circ Rec(t, D \cup \{t\}) \circ \Pi(t, v, D \cup \{t\}) \quad \text{(III)}$$

$$Rec(u, D) = \{\varepsilon\} \cup \Pi(u, u, D) \circ Rec(u, D) \quad \text{(IV)}$$

**Лемма 5.** *$\forall p \in \Pi(u, v, D)$ начало пути $beg(p) = u$, конец пути $end(p) = v$.*

*Доказательство.* Очевидно.

**Лемма 6.** *$\forall u, v, D$ верно, что $\varepsilon \notin \Pi(u, v, D)$.*

*Доказательство.* Очевидно, поскольку согласно правилам (I – III), любой путь $p \in \Pi(u, v, D)$ начинается с ребра $(u, t)$.

**Лемма 7.** *Пусть дан путь $p = s_1 \to \ldots \to s_n$ такой, что $\forall i > 1, s_i \neq s_1$.*

Тогда $p \in Rec(s_1, D) \circ \Pi(s_1, s_n, D) \Leftrightarrow p \in \Pi(s_1, s_n, D)$.

*Доказательство. Достаточность.* Очевидно, поскольку $\varepsilon \in Rec(s_1, D)$.

*Необходимость.* Раскроем определение $Rec(s_1, D)$.

$$Rec(s_1, D) \circ \Pi(s_1, s_n, D) = \begin{cases} \Pi(s_1, s_n, D) \\ \Pi(s_1, s_1, D) \circ Rec(s_1, D) \circ \Pi(s_1, s_n, D) \end{cases} \quad (15)$$

Согласно леммам 5 , 6, $\forall p \in \Pi(s_1, s_1, D) \circ Rec(s_1, D) \circ \Pi(s_1, s_n, D)$ вершина $s_1$ встречается в $p$ как минимум два раза, но по условию, $s_1$ встречается только один раз. Следовательно, единственная возможность — $p \in \Pi(s_1, s_n, D)$.

**Лемма 8.** *Пусть даны множество $D$ такое, что $D \subseteq RV$, и путь $p = s_1 \rightarrow \ldots \rightarrow s_n$, удовлетворяющий условиям:*

4. длина $p \geq 1$;
5. $s_n = exit$ или $s_n \in D$;
6. $\forall i < n, \quad s_i \notin D \backslash \{s_1\}$.

Тогда верно следующее.

- Если $s_1 \notin RV$, то $p \in \Pi(s_1, s_n, D)$.
- Если $s_1 \in RV$, то $p \in Rec(s_1, D) \circ \Pi(s_1, s_n, D)$.

*Доказательство.* Индукция #1 по длине пути $p$. База: длина пути равна 1. Тогда $p = s_1 \rightarrow s_n \in \{(s_1, s_n)\} \subseteq \Pi(s_1, s_n, D)$. Если $s_1 \in RV$, то $p \in Rec(s_1, D) \circ \Pi(s_1, s_n, D)$, поскольку $\varepsilon \in Rec(s_1, D)$.

Индукционный переход.

- Пусть $p = s_1 \rightarrow v \rightarrow \ldots \rightarrow s_n$, и предположим, что вершина $s_1$ встречается в пути $p$ ровно 1 раз.
  - Если вершина $v \notin RV$, то по И.П. #1, путь $v \rightarrow \ldots \rightarrow s_n \in \Pi(v, s_n, D)$, и, по правилу (II), имеем $p \in \Pi(s_1, s_n, D)$, так как $v \neq s_n$.
  - Пусть вершина $v \in RV$. Тогда по условию (3) можно сделать вывод, что $v \notin D$. Поскольку для всех вершин $u$, кроме последней, в пути $v \rightarrow \ldots \rightarrow s_n$ выполнено $u \notin D \cup \{v\} \backslash \{v\}$, так как среди них нет $s_1$, то по И.П. #1, верно, что $v \rightarrow \ldots \rightarrow s_n \in Rec(v, D \cup \{v\}) \circ \Pi(v, s_n, D \cup \{v\})$, а значит, по правилу (III), $p \in \Pi(s_1, s_n, D)$, так как $v \neq s_n$.

  Если вершина $s_1 \in RV$, то $p \in Rec(s_1, D) \circ \Pi(s_1, s_n, D)$, поскольку $\varepsilon \in Rec(s_1, D)$.
- Пусть $p = s_1 \rightarrow v \rightarrow \ldots \rightarrow s_1 \rightarrow \ldots \rightarrow s_n$ и $s_1 \in RV$. Обозначим $p \equiv q \circ p'$, где $q \equiv s_1 \rightarrow \ldots \rightarrow s_1$, а $p' \equiv s_1 \rightarrow \ldots \rightarrow s_n$ такой, что вершина $s_1$ встречается в пути $p'$ ровно один раз, если $s_n \neq s_1$, или ровно два раза, если $s_n = s_1$. Если $s_1 \neq s_n$, то по И.П. #1, $p' \in Rec(s_1, D) \circ \Pi(s_1, s_n, D)$, откуда по лемме 7, $p' \in \Pi(s_1, s_n, D)$. Если $s_n = s_1$, то рассмотрим подробнее путь $p'$.
  - $p' \equiv s_1 \rightarrow s_1$, откуда $q' \in \{(s_1, s_1)\} \subseteq \Pi(s_1, s_1, D)$.
  - $p' \equiv s_1 \rightarrow v_1 \rightarrow \ldots \rightarrow v_m \rightarrow s_1$ такой, что $v_1 \notin RV$. Так как $\forall 1 \leq i \leq m$ верно, что $v_i \notin D$, то по И.П. #1, путь $v_1 \rightarrow \ldots \rightarrow v_m \rightarrow s_1 \in$ $\Pi(v_1, s_1, D)$, а, по правилу (II), получаем, что $p' \in \Pi(s_1, s_1, D)$, так как $v_1 \neq s_1$.

- $p' \equiv s_1 \to v_1 \to \ldots \to v_m \to s_1$ такой , что $v_1 \in RV$. Так как $\forall 1 \leq i \leq m$ верно, что $v_i \notin D \cup \{v_1\} \backslash \{v_1\}$, то по И.П. #1, путь $v_1 \to \ldots \to v_m \to s_1 \in Rec(s_1, D \cup \{v_1\}) \circ \Pi(v_1, s_n, D \cup \{v_1\})$, а, по правилу (II), получаем, что $p' \in \Pi(s_1, s_1, D)$, так как $v_1 \neq s_1$.

Осталось показать, что $q \in Rec(s_1, D)$. Индукция #2 по длине q. База $q \equiv \varepsilon$ — очевидно.

Переход. Пусть $q \equiv q' \circ q''$, где $q' \equiv s_1 \to \ldots \to s_1$ и содержит только два вхождения $s_1$, а $q'' \equiv s_1 \to \ldots \to s_1$ является остатком пути $q$. Но по И.П. #2, получаем, что $q'' \in Rec(s_1, D)$. Аналогично получаем, что $q' \in \Pi(s_1, s_1, D)$. Тогда согласно правилу (IV), получаем, что $q \in Rec(s_1, D)$. Но тогда $p \equiv q \circ p' \in Rec(s_1, D) \circ \Pi(s_1, s_n, D)$.

- Пусть $p = s_1 \to v \to \ldots \to s_1 \to \ldots \to s_n$ и $s_1 \notin RV$. Покажем, что $v \neq s_n$. Пусть это не так и $v = s_n$. Тогда единственная возможность — $s_n \in D$, потому что из вершины $exit$ не исходит ребер. Но в таком случае не выполняется условие (3) для пути $p$ в вершине $v$, потому что $D \backslash \{s_1\} = D$. Пришли к противоречию, значит, $v \neq s_n$. Если вершина $v \notin RV$, то по И.П. #1, получаем, что $v \to \ldots \to s_1 \to \ldots \to s_n \in \Pi(v, s_n, D)$ и, по правилу (II), $p \in \Pi(s_1, s_n, D)$. Если вершина $v \in RV$, то по И.П. #1, для $D := D \cup \{v\}$ получаем, что $v \to \ldots \to s_1 \to \ldots \to s_n \in Rec(v, D \cup \{v\}) \circ \Pi(v, s_n, D \cup \{v\})$ и, по правилу (III), $p \in \Pi(s_1, s_n, D)$.

**Теорема 20.** (Пути в графе потока управления). $p = start \to \ldots \to exit$ *— путь в графе* $G$ *тогда и только тогда, когда* $p \in \Pi(start, exit, \varnothing)$.

*Доказательство. Достаточность.* Очевидно.

*Необходимость.*

Воспользуемся условиями леммы 8:

- так как $start \neq exit$, то длина пути $p \geq 1$;
- $\forall i < n,\ s_i \notin \varnothing \backslash \{start\}$;
- последняя вершина равна $exit$;
- $start \notin RV$.

Тогда получаем, что $p \in \Pi(start, exit, \varnothing)$.